\documentclass[12pt]{article}
\usepackage{amssymb}
\begin{document}
%\language=0
%\large
 \begin{center}
{\Large\bf SIMULTANEOUSLY DISSIPATIVE OPERATORS\\
AND THE INFINITESIMAL MOORE EFFECT\\[1.5mm]
IN INTERVAL SPACES}\\[12mm]
\end{center}
\begin{center}
A.N.Gorban$^\dag$\footnote{Since 2004 at the Department of Mathematics, University of
Leicester, UK;  \texttt{ag153@le.ac.uk}},
Yu.I.Shokin$^\ddag$, V.I.Verbitskii$^\dag$\\[4mm]
$^\dag${\it Krasnoyarsk Computing Center\\
Russian Academy of Sciences, Siberian Branch\\
Krasnoyarsk-36, 660036, Russian Federation}\\[2mm]
$^\ddag${\it Institute of Computational Technologies\\
Russian Academy of Sciences, Siberian Branch\\
Novosibirsk-90, 630090, Russian Federation}\\[4mm]
\end{center}

\begin{abstract}
One of shortcomings of stepwise interval methods is the following. The intervals
determining the solution of a system are often expanded in the course of time
irrespective of the method and step used (the {\em Moore effect}). We introduce the
notion of general {\em interval spaces} and study the infinitesimal Moore effect (IME) in
these spaces. We obtain the local conditions of absence of the IME in terms of Jacobi
matrices field. The relation between the absence of IME and simultaneous dissipativity of
the Jacobi matrices is established. We study simultaneously dissipative operators in
$\Bbb{R}^n$. A linear operator $A$ is {\em dissipative} with respect to a norm $\|...\|$
if $\| \exp (At) \| \leq 1$ at all $t \geq 0$. For each norm, the dissipative operator
form a closed convex cone. An operator $A$ is {\em stable dissipative} if it belongs to
the interior of this cone. The family of linear operators $\{A_\alpha \}$ is called {\em
simultaneously dissipative}, if there exists a norm with respect to which all the
operators are dissipative. We studied general properties of such families. For example,
let the family $\{A_\alpha \}$ be finite and generate a nilpotent Lee algebra and let for
each $A_\alpha$ there exist a norm with respect to which it is dissipative. Then
$\{A_\alpha \}$ is simultaneously dissipative. Let the family $\{A_\alpha \}$ be compact
and generate solvable Lee algebra, and let the spectrum of each operator $A_\alpha $ lie
in the open left half-plane. Then $\{A_\alpha \}$ is simultaneously stable dissipative,
i.e. there exists a norm with respect to which all $A_\alpha $ are stable dissipative. We
study the conditions of simultaneous dissipativity of the matrices of rank one and
discussed their application to equations of {\em mass action law} kinetics.
\end{abstract}

\section{INTRODUCTION}
For solving systems of ordinary differential equations different classes of numerical
methods with guaranteed error estimation including interval methods are used. In solving
a system by an interval method the approximate solution at any considered moment of time
$t$ represents a set (called interval) containing the exact solution at the moment $t$.
The detailed account of interval methods can be found in monographs by R.Moore [1] and
S.A.Kalmykov, Yu.I.Shokin, Z.Kh.Yuldashev [2].

As a rule, all kinds of rectangular parallelepipeds with sides
parallel to coordinate axes [1,2] are used as intervals, less
frequently -- ellipsoids [3], balls of fixed norm [4,5] etc.

One of shortcomings of stepwise interval methods is the following.
The intervals determining the solution of a system are often
expanded in the course of time irrespective of the method and step
used. The simplest example of strong expansion of intervals during
a short time, belonging to R.Moore, is given in [1]. The
phenomenon of interval expansion, called the Moore sweep effect
(or simply the Moore effect), essentially decreases the efficiency
of interval methods. Earlier the Moore effect was investigated
only for some particular systems and particular intervals [1].

In the present work the notions of the interval and the Moore
effect are formalized and the Moore effect is studied for
autonomous systems on positively invariant convex compact.

Formally, one can get rid of the interval expansion for any globally stable system (i.e.
such a system, any solution of which is stable according to Lyapunov). To demonstrate
that, let consider a smooth autonomous system: $$ \frac{d x}{d t}=f(x) \eqno{(1)} $$ on
the positively invariant compact $B \subset \Bbb{R}^n$. Construct a metric $\rho$ on the
set $B$, assuming for any $x \in B, y \in B$: $$ \rho(x, y)=\sup_{t\ge 0} \Vert
x(t)-y(t)\Vert, $$ where \ $x(t),\ y(t)$ \ are the solutions of the system (1) with the
initial conditions $x(0)=~x,$ $y(0)=~y$. This metric is contracting for (1), i.e. for any
pair $x(t)$, $y(t)$ of the solutions of (1) with the initial conditions in $B$
$$ \rho(x(t), y(t))\le \rho(x(s), y(s)) \ \mbox{at} \ t\ge s. $$
The metric $\rho$ is topologically equivalent to norm if and only
if the system (1) is globally stable in $B$. If one considers as
intervals all balls of the metric $\rho$, then in a definite sense
the Moore effect is absent. That is, there is no interval
expansion when constructing the exact interval solution with any
step $h>0$. The exact interval solution of $X(t)$ is defined in
the following way: $X(0)=X_0$, where $X_0$ is the initial interval
with the centre at the point $x(0)=x_0$;\ $X((n+1)h)$ is the
minimal interval with the centre at the point $x((n+1)h)$
containing $T_hX(nh)$, where $T_t$ is the transformation of the
phase flow of (1) during the time $t\ge0$. Indeed, the radius
$X((n+1)h)$ does not exceed the radius $X(nh)$ at any $n$.

If the system is not globally stable, then metric is not
topologically equivalent to the norm. It means that small, in
usual sense, intervals became large in the metric $\rho$. This
circumstance makes one refuse from consideration of similar
metrics. Moreover, if the system (1) is absolutely unstable (for
example, a system with mixing), then there is no reasonable way to
get rid of the Moore effect.

The decsribed method of elimination the Moore effect for globally stable system is
non-constructive. This can be demonstrated as follows: for constructing the contracting
metric $\rho$ one must know all exact solutions of the system (1). But then it is
unreasonable to solve the system numerically. We must have constructively verifiable
conditions of absence of the Moore

is what we deal with in the present paper. The conditions of
absence of the Moore effect are of local character and formulated
in terms of Jacobi matrices of the system. Except that the causes
of frequent appearance of the Moore effect will be pointed out.

\section{Interval spaces and the Moore effect}
\subsection{Interval Spaces}

Before starting to study the Moore effect, it is necessary to
define what we mean by intervals. Generalizing known
constructions, give the following definition.

{\bf Definition 1}. We call the family $\Bbb{J}$ of convex
compacts in $\Bbb{R}^n$ {\it the interval space} (and its elements
-- {\it intervals}), if it satisfies the following conditions:

a) $\Bbb{J}$ is closed with respect to multiplication by
non-negative scalars: $$ \mbox{if} \ W \in \Bbb{J},\  \alpha \ge
0,\ \mbox{then} \ \alpha W=\{\alpha x \mid x \in W\} \in \Bbb{J};
$$

b) $\Bbb{J}$ is closed with respect to intersection: $$ \mbox{if}
\ W_1 \in \Bbb{J},\ W_2 \in \Bbb{J}, \ \mbox{then} \ W_1\cap W_2
\in \Bbb{J}; $$

c) $\Bbb{J}$ is closed according to Hausdorff (i.e. in the
Hausdorff metric);

d) if $W \in \Bbb{J},\ W\neq\{0\}$, then $0 \in riW$.

Remind [6] that the Hausdorff metric on the set of all compacts in
$\Bbb{R}^n$ is introduced as follows: $$ \rho_H(x, y)=\max
\{\max_{x \in X}\ \min_{y \in Y}\ \Vert x-y \Vert ,
\
\max_{y \in Y}\ \min_{x \in X}\ \Vert x-y \Vert\}, $$ where $x, y$
are the compacts in $\Bbb{R}^n$,\ \ $\Vert . \Vert $ is a fixed
norm in $\Bbb{R}^n$. All Hausdorff metrics in $\Bbb{R}^n$ are
equivalent.

Further on by $\lim_{H i \to \infty} W_i$ we denote the Hausdorff
limit of the sequence $ \{W_i\}_{i=1}^{+\infty}$\ at \ $i \to
\infty$.

Give several examples of interval spaces.

{\bf Example 1}. $\Bbb{J}$ is the set of all convex compacts
symmetric with respect to $0$. It satisfies all the properties
from a) to d).

{\bf Example 2}. $\Bbb{J}$ is the set of all symmetric with
respect to $0$ rectangular parallelepipeds (including
non-singular), i.e. sets of the form $$ \{x=(x_1,\ldots,x_n) \in
\Bbb{R}^n:
 \  \vert x_k \vert \leq a_k
 \  (k=1,\ldots,n)\},
$$ where $ a_k \geq 0  \  (k=1,\ldots,n).$ It satisfies the
properties a),  b) and d).

Let now ${\{W_i\}}_{i=1}^\infty \subset \Bbb{J},\  \lim_{H i \to
\infty} W_i=W$ \  with \  $a_k^{(i)}$ being $a_k$, corresponding
to $W_i$.

If $\rho_H(W_i,W) < \varepsilon, \  \mbox{then} \  W \subset
W_i+P_\varepsilon, \  W_i \subset W+P_\varepsilon, \  \mbox{where}
\  P_\varepsilon =\{x \in \Bbb{R}^n: \  \vert x_k\vert \leq
\varepsilon \  (k=1,\ldots,n)\}$ \  (here a norm in the definition
of the Hausdorff metric is the $l^\infty$-norm). Then for any $x
\in W$ $$ \vert x_k \vert \leq a_k^{(i)}+\varepsilon \ \
(k=1,\ldots,n) $$ is true and for any $x \in W_i$ $$ \vert x_k
\vert \leq \underline{\lim}_{i \to \infty} a_k^{(i)}; $$ $$
\lim_{i \to \infty} a_k^{(i)} \leq \overline{\lim}_{i \to \infty}
a_k^{(i)}, $$ \ i.e. there exist the limits $$
 \tilde a_k=\lim_{i \to \infty} a_k^{(i)} \ \  (k=1,\ldots,n).
$$ If $x \in W$,\ then $$ \vert x_k \vert \leq \lim_{i \to \infty}
a_k^{(i)} \ \ (k=1,\ldots,n). \eqno {(2)} $$ Let $b_i=\min_{1 \le
k} (a_k^{(i)} / \tilde a_k)$. If for some $x \in \Bbb{R}^n$ the
inequalities (2) are satisfied, then $\lim_{i \to \infty} b_i=1$.

Obviously, $x^{(i)}=b_ix\in W_i$, i.e. there exists such a
subsequence of $\{x^{(i)}\}_{i=1}^\infty$ that $x^{(i)} \in W_i$,
$ x=\lim_{i \to \infty} x_{(i)}$. Hence $W=\{x \in \Bbb{R}^n\ :\
\vert x_k \vert \le \tilde a_k \  (k=1,\ldots,n)\}$, i.e. $W \in
\Bbb{J}$ and the property c) is also satisfied.

In constructing interval methods of solving different problems it
is, as a rule, the considered interval space that is made use of
[1,2].

{\bf Example 3}. Let $\Vert . \Vert $ be a norm in $\Bbb{R}^n$, $\mathbf{B}_r=\{x\in
\Bbb{R}^n: \Vert x \Vert \le r \}$, where $r\ge 0$. Let $$
\Bbb{J}=\{\mathbf{B}\mathbf{}_r \vert \ r\geq 0 \},
$$ i.e. J is the set of all closed balls (further on we omit the word "closed") of the
norm. All the properties from a) to d) are satisfied. These interval spaces are used, for
example, in [4,5].

{\bf Example 4}. The construction of example 3 can be generalized as follows. Let ${\Vert
. \Vert}_1, \ldots ,{\Vert . \Vert}_m$ be the finite set of norms in $\Bbb{R}^n,$ $$
\mathbf{B}_{r_k}^{(k)}=\{x \in \Bbb{R}^n: {\  \Vert x \Vert}_k \leq r_k \ \  (k=1,\ldots
, m)\} $$ where $r_k \geq 0 \ \  (k=1, \ldots, m)$ \quad and $$ W_{r_1, \ldots,
r_m}=\bigcap_{1 \leq k \leq m} \mathbf{B}_{r_k}^{(k)}. $$ Let $\Bbb{J}=\{W_{r_1, \ldots ,
r_m}: \quad r_k \geq 0 \ \ (k=1, \ldots, m)\}.$ \  Obviously,   J possess the properties
a),  b), and d).

Note that the same element of $\Bbb{J}$ can be associated with different sets of
$\{r_k\}$. To demonstrate that, let ${m=2}, \sup_{x \ne 0} ({\Vert x \Vert}_2 / {\Vert x
\Vert}_1)=C.$ Then $\mathbf{B}_1^{(1)}=W_{1,C'}$, where $C'$ is any number not less than
$C$. Also, even if one of $r_k$ is equal to $0$, then $$ W_{r_1, \ldots, r_m}=\{0\}. $$

To each compact $W \subset \Bbb{R}^n$ can be juxtaposed the set $$
{\{r_k(W)\}}_{k=1}^m: \ \ r_k(W)=\max_{x \in W} {\Vert x \Vert}_k
\ \  (k=1, \ldots, m). $$ If $W \in \Bbb{J}$ \  then \ $W=W_{r_1}
\cap \ldots \cap W_{r_m}.$

Let now the sequence ${\{W_i\}}_{i=1}^\infty $ converge according to Hausdorff to the
compact $W$, with $W_i \in \Bbb{J}$ for all $i$. Similarly to example 2, from the
inclusions $$ W_i \subset W+\mathbf{B}_{\varepsilon }^{(k)}, \ \  W\subset
W_i+\mathbf{B}_{\varepsilon }^{(k)} $$ satisfied for each $\varepsilon >0$ for all
$i>i_0(\varepsilon)$ derive the existence of the limits: $$ \tilde r_k=\lim_{i \to
\infty} r_k(W_i) \ \ (k=1,\ldots, m) $$ and conclude that $$ W=W_{\tilde r_1,\ldots
,\tilde r_m}, $$ i.e. $W \in \Bbb{J}$, and the property \  c) \ is satisfied.

{\bf Example 5}. Let $Q$ be a compact convex body without symmetry
centre (for instance, a triangle in $\Bbb{R}^2$), $0 \in int Q$.
Assume $$ \Bbb{J}=\{\alpha Q: \  \alpha \geq 0\}. $$ J possesses
the properties from a) to d).

{\bf Remark 1}. Example 5 can be generalized. For this purpose it
is necessary to consider compact convex bodies $Q_1, \ldots, Q_m,$
the interior of each of them contains $0$, and to take as
$\Bbb{J}$ a family of all the sets of the form $\bigcap_{1 \leq k
\leq m} {\alpha }_kQ_k$ where ${\alpha }_k~\geq~0\ \  (k=1,\ldots
, m).$

\subsection {Dissipative Operators}

In this section the properties of the operators dissipative with
respect to compact are studied. First, let remind some notations.

The affine envelope of the convex set $W$ is denoted by $AffW$,
the relative interior $W$ (the interior of $W$ in $AffW$) is
denoted by $riW$, the relative boundary of $W$ (the boundary of
$W$ in $AffW$) is denoted by $r\partial W$. For the boundary of
the set $X$ we use the notation $\partial X$, $int X$ -- for the
interior of $X$, $co X$ -- for the convex envelope of $X$. By the
sum of the sets of $X$ and $Y$ from ${\Bbb{R}}^n$ we mean the set
$\{x+y: \  x \in X, y \in Y\}$, by I -- the unit operator.

Let introduce a new notion.

{\bf Definition 2}. The linear operator $A$ in the space
${\Bbb{R}}^n$ is called {\it dissipative with respect to the
family of sets} $\{W_\nu \} \subset {\Bbb{R}}^n$ if every set
$W_\nu$ is positively invariant with respect to the system $$
\frac {d x}{d t}=Ax. \eqno{(3)} $$

In other words, every $W_\nu $ is invariant with respect to the
semi-group of the operators $\exp (At) \  (t \geq 0).$

Below we consider operators dissipative with respect to families
of convex compacts. In particular, the operator is dissipative
with respect to the families of all balls of some norm (for this,
dissipativity with respect to only one ball is sufficient) if and
only if $\Vert \exp (At) \Vert \leq 1$ at all $t \geq 0$. Thus, in
this case we come to the known definition of dissipativity with
respect to the norm {[7]}.

The set of all operators dissipative with respect to $\{W_{\nu
}\}$ is denoted by $K(\{W_{\nu }\})$.

{\bf Remark 2}. If an operator is dissipative with respect to the
family of compacts and the interior of at least one of them is not
empty, then it is dissipative with respect to some norm.

Indeed, any symmetric with respect to $0$ compact convex body is a
ball of some norm (see, for example, [7]). Choose as a ball the
following set: $$ S=co \{W \cup (-W)\} \eqno {(4)} $$ where $W$ is
any set of the considered family of $\{W_{\nu }\}$, for which $int
W \ne \emptyset$.

However, if $W$ is a compact and the operator is dissipative with
respect to the norm whose ball is $S$ (4), then it does not yet
mean that the operator is dissipative with respect to $W$ (see
also example 8).

{\bf Remark 3}. From the invariance of a family of compacts with
respect to the linear operator follows the invariance of the
Hausdorff closure (i.e. closure in the Hausdorff metric) of this
family. Therefore from dissipativity of the operator with respect
to the family of compacts follows the dissipativity with respect
to Hausdorff closure of this family.

Let $W$ be a convex compact in ${\Bbb{R}}^n$ with $0 \in riW$. In
this case $AffW$ is a linear subspace, and if the operator $A$ is
dissipative with respect to $W$, then $AffW$ is invariant with
respect to $A$. Introduce the following functional on the subspace
$L(W)$ of the space $L({\Bbb{R}}^n)$ (of linear operators in
${\Bbb{R}}^n$), consisting of the operators, with respect to which
$AffW$ is invariant: $$ {\mu }_W (A)=\sup_{x \in W} {\mu }_W (Ax).
\eqno{(5)} $$ Here ${\mu }_W$ is the Minkovski functional of the
set $W$ (defined, for example, in [8]) in the subspace $AffW$.

It is easy to see that $A \in K(W)$ if and only if $$ {\mu }_W
(\exp (At)) \leq 1 $$ for all $t \geq 0$.

In particular, the operator $A \in L(W)$ is {\it strongly
dissipative with respect to convex compact} $W$ if exists such
$\varepsilon >0$ that ${\mu }_W (\exp (At)) \leq \exp
(-\varepsilon t)$ at all $t \geq 0$. In general, the operator $A$
is strongly dissipative with respect to convex compact $W$ if and
only if $A+ \varepsilon I \in K(W)$ for some $\varepsilon >0$.

If $W$ is a ball of the norm $\Vert . \Vert$, then strong
dissipativity with respect to $W$ means the existence of such
$\varepsilon >0$ that $\Vert \exp (At) \Vert \leq \exp
(-\varepsilon t)$ for all $t \geq 0$. We come to the definition of
{\it stable dissipativity with respect to the norm} [11, 12, 19].

Introduce in $L({\Bbb{R}}^n)$ the following functional: $$ {\gamma
}_W (A) = \lim_{h \to +0} \frac {{\mu }_W (I+hA)-1}{h} $$ In the
case, when $W$ is a ball of some norm (i.e. ${\mu }_W$ is a norm),
arrive at the known definition of the logarithmic Lozinsky norm
[9, 10].

{\bf Lemma 1}. The operator $A \in L({\Bbb{R}}^n)$ is dissipative
(strongly dissipative) with respect to $W$, if and only if the
inequality ${\gamma }_W (A) \leq 0 \ \ ({\gamma }_W (A) < 0)$ is
satisfied.

{\bf Proof}. {\it Sufficiency}. The following inequality is
obtained in [9] $$ \Vert \exp (At) \Vert \leq \exp (\gamma (A)t)
$$ where $\gamma (A)$ is the Lozinsky norm of the operator $A$,
corresponding to the norm $\Vert . \Vert$. By literal repetition
of the reasonings from [9] (with a substitution of the norm by
Minkovski functional), one can obtain the inequaliny $$ {\mu }_W
(\exp (At)) \leq \exp ({\gamma }_W (A)t) $$ for all $t \geq 0$,
from which immediatelly follows the sufficiency.

{\it Necessity}. Evidently, $$ {\mu }_W (\exp (At))={\mu }_W
(I+At)+o(t) \  (t \to 0). $$ Therefore, $$ {\gamma }_W (A)=\lim_{h
\to +0} \frac {{\mu }_W (e^{Ah})-1}{h}. $$ Let $\varepsilon \geq
0$. If ${\mu }_W (\exp (At)) \leq \exp (-\varepsilon t)$ at all $t
\geq 0$, then $$ {\gamma }_W (A) \leq \lim_{h \to +0} \frac {\exp
(-\varepsilon h)-1} {h}= -\varepsilon , $$ which proves the
necessity. The lemma is proved.

Assign a relatively open convex cone $Q_x(W)$ to every point $x
\in r\partial W$ according to the rule: $y \in Q_x(W)$ if and only
if there exists such $ \varepsilon > 0$ that $$ x+\varepsilon y\in
riW. $$

{\bf Lemma 2}. For strong dissipativity of $A$ with respect to
convex compact $W$ it is necessary and sufficient that for every
point $x \in r\partial W$ the inclusion $$ Ax \in Q_x (W) $$ be
true. For dissipativity of $A$ with respect to $W$ it is necessary
and sufficient that for every point of $X \in r\partial W$ the
inclusion $$ Ax \in \overline {Q_x (W)} $$ be true.

{\bf Proof}. Note that the operator $A$ is strongly dissipative
with respect to $W$ if and only if there exists such $t_0 >0$ that
${\mu }_W (I+At_0) < 1$. Indeed, the existence of such $t_0$ for a
strongly dissipative operator follows immediately from the
negativeness of ${\gamma }_W (A)$. Conversely, if ${\mu }_W
(I+At_0) < 1$, then there exists such ${\varepsilon } > 0$ that
${\mu }_W (I+(A+\varepsilon I)t_0) < 1$. But then ${\gamma }_W
(A+\varepsilon I) \leq 0$, the operator $(A+\varepsilon I)$ is
dissipative. It means that $A$ is strongly dissipative.

If the operator $A$ is strongly dissipative with respect to $W$,
then, according to the above, for each $x \in r\partial W$ there
exists such $t_x > 0$ that $(I+t_xA)x \in riW$. It means that the
vector $Ax$ belongs to to the cone $Q_x(W)$.

Conversely, let the latter condition be satisfied. According to
the hypothesis of the theorem and convexity of $W$, for each $x
\in r\partial W$ there exists the only positive number $s=s(x)$
such that $(I+sA)x \in r\partial W$. Show that $s_0=\inf_{x \in
r\partial W} s(x) > 0$. Let it be not so. Then there exists such a
subsequence ${\{x_n\}}_{n=1}^{+\infty }$ that $\lim_{n \to \infty}
s(x_n)=0$. Choose from $\{x_n\}$ a converging subsequence
$\{{x_n}'\}$. Let $\tilde x=\lim_{n \to \infty } {x_n}'$. For
every $n \in N$ and for every $\varepsilon > 0$ $$
[I+(s({x_n}')+\varepsilon )A] {x_n}' \notin W. $$ Passing to the
limit, obtain $$ (I+\varepsilon A) \tilde x \notin ri W $$ which
contradicts the hypothesis of the theorem.

Thus, $s_0 > 0$. For any $t_0 \in (0;s_0)$ is true ${\mu }_W
(I+At_0) < 1$, i.e. the operator $A$ is strongly dissipative.

If $A \in K(W)$, then for any $\varepsilon > 0$ we have
$AX-\varepsilon x \in Q_x(W)$ (for any $x \in r\partial W$), i.e.
$Ax \in \bar Q_x$. Conversely, if $Ax \in \bar Q_x$, then
$Ax-\varepsilon x \in Qx$ at any $\varepsilon > 0$, and $A$
represents a limit point of the family of dissipative operators,
i.e. $A \in K(W)$. The lemma is proved.

{\bf Remark 4}. Immediately from the Krein-Milman theorem [8]
follows that it is sufficient to require from the lemma conditions
that inclusions be satisfied not for all points $x \in r\partial
W$, but for extremal points of $W$ only. In particular, if $W$ is
a polyhedron, then it is sufficient to test its vertices only.
Thus, to elucidate the question about dissipativity (strong
dissipativity) of the operator with respect to the polyhedron, one
should test only the fulfilment of finite number of linear
inequalities.

{\bf Remark 5}. In the proof lemma 2 we have used the obvious
fact: the closure of the set $K(W)$.

One more fact follows directly from lemma 2.

{\bf Lemma 3}. The set $K(W)$ is a closed convex cone. The cone of
all strongly dissipative with respect to $W$ operators coincides
with $ri K(W)$ and with $\bigcup_{\varepsilon > 0}
(K(W)-\varepsilon I)$. If $\{W_\nu \}$ is a family of convex
compacts with $0 \in ri W_\nu $ for all $\nu $, then $K(\{W_\nu
\})$ is a closed convex cone.

{\bf Remark 6}. If $int W=\emptyset$, then $int K(W)=\emptyset$.
Indeed, if $AffW$ is invariant with respect to the operator $A_1$,
then $A+\varepsilon A_1 \notin K(W)$ at $\varepsilon \ne 0$. If
$int W \ne \emptyset$, then $int K(W)$ is also non-empty and
coincides with $ri K(W)$.

{\bf Definition 4}. The operator $A \in K(W)$ is called {\it
stable (or roughly) dissipative with respect to} $W$, if $A \in
int K(W)$.

Definition 4 generalize the definition of the stable dissipativity
with respect to the norm [11, 19].

Pass to the consideration of operators dissipative with respect to
interval spaces. Let find out for which interval spaces $\Bbb{J}$
the interior of the cone $K(\Bbb{J}) $ is not empty.

Let $V$ be a set of all compact convex bodies in $\Bbb{R}_n$. Fix
some norm $\Vert . \Vert$ in $\Bbb{R}^n$ and assume $$
d(W)=\min_{x \in \partial W} {\Vert x \Vert}. $$

{\bf Lemma 4}. The function $d(W)$ is continuous according to
Hausdorff on the set $V$.

{\bf Proof}. First note that if $X \in V, \  Y \in V$, then ${\rho
}_H (\partial X, \partial Y) \leq {\rho }_H (X, Y)$. Indeed, let
${\rho }_H (X, Y) \leq \varepsilon $. Then $X \subset
Y+S_{\varepsilon }$ where $S_{\varepsilon }=\{x \in \Bbb{R}^n: \
\Vert x \Vert \leq \varepsilon \}$. Let, further on, there exists
such $y_0 \in (\partial Y) \cap X$ that $y_0 \notin S_{\varepsilon
} + \partial X$. Construct at the point $y_0$ a tangent hyperplane
$L$ to $Y$. Let $l$ be the direction of the external normal to
$\partial Y$ at the point $y_0$ orthogonal to $L$. Draw a ray from
the point $y_0$ in the direction of $l$ to the point $x_0$ of
crossing with $\partial X$. Construct such a ball $S$ of the norm
$\Vert . \Vert$ with the centre at the point $x$ that $y_0 \in
\partial S$. The radius of $S$ is larger than $\varepsilon $ and
$S \cap Y = \{y_0\}$. Thus, if one constructs a ball $S'\subset S$
of the radius $\varepsilon $ with the centre at $x_0$, then $$ S'
\cap Y = 0. $$ But then $x_0 \notin Y+S_{\varepsilon }$, i.e. $X
\not\subset Y+S_{\varepsilon }$ what is contrary to the
assumption.

The existence of such $y_0 \subset \partial Y$ that $y_0 \notin X
\cup (\partial X +S_{\varepsilon })$ is also impossible, since
then $y_0 \notin X+S_{\varepsilon }$, i.e. $Y \not\subset
X+S_{\varepsilon }$. Consequently, $\partial Y \subset
S_{\varepsilon }+\partial X$, and that means $d(X) \leq
d(Y)+\varepsilon $. Similarly, $d(Y) \leq d(X)+\varepsilon $. It
means that $\vert d(X)-d(Y) \vert \leq \varepsilon $, and the
function $d(W)$ is continuous on $V$. The lemma is proved.

{\bf Lemma 5}. For non-emptiness of $int K(\Bbb{J})$ it is
necessary and sufficient for all the elements of the interval
space $\Bbb{J}$, exept $\{0\}$, to posess non-empty interior.

{\bf Proof}. {\it Necessity}. Follows immediately from remark 6.

{\it Sufficiency}. Show that under the conditions of the theorem
the inclusion $$ -I \in int K(\Bbb{J}) \eqno {(6)} $$ takes place.

To each point $x \ (\Vert x \Vert =1)$ we assign the set $W_x$
according to the rule: $$ W_x = \bigcap_{W \ni x, W \in \Bbb{J}}
W. $$

According to the conditions b) and c) from definition 1, $W_x \in
\Bbb{J}$. The set $$ \tilde W= \overline {\bigcup_{\Vert x \Vert
=1} W_x} $$ is compact. Indeed, $\tilde W$ is contained in any
element of $\Bbb{J}$ containing unit ball of the norm $\Vert .
\Vert$; such an element exists due to non-emptiness of the
interior of all intervals (exept $\{0\}$) and the property a) from
definition 1. Note that Hausdorff closure of the family $\{W_x:\
\Vert x \Vert =1\}$ represents a compact in the Hausdorff metric,
contained in $\Bbb{J}$ (this follows from compactness according to
Hausdorff of the family of all compact subsets of the compact
[6]). From the property a) (definition 1) follows (by virtue of
lemma 4) the existence of such $\varepsilon >0$ that $d(W_x) \geq
\varepsilon $ for all such $x$ that $\Vert x \Vert =1$ (indeed,
$d(W_x) >0$, since $0 \in int W_x$).

Thus, there exists such $\varepsilon >0$ that for all $x\ (\Vert x
\Vert =1)$ the inclusion $$ Ax \in int W_x $$ is true if $\Vert A
\Vert < \varepsilon $.

In other words, $Ax-x \in Q_x (W_x)$ if $\Vert A \Vert <
\varepsilon , \  \Vert x \Vert =1$ (see lemma 1). The more so, as
$Ax-x \in Q_x(W)$ for all $W \in \Bbb{J} \ (W \ni x, \Vert A \Vert
< \varepsilon )$ at all such $x$ that $\Vert x \Vert =1$. But then
$Ax-x \in Q_{\alpha x}(\alpha W)$ for all $\alpha >0, \Vert A
\Vert < \varepsilon$. Hence, $A-I \in K(\Bbb{J})$, i.e. (6) is
satisfied. The lemma is proved.

Thus, we have shown that under the conditions of lemma 5
$K(\Bbb{J})$ is a convex solid cone.

{\bf Definition 5}. The operator is {\it stable dissipative with
respect to the interval space} $\Bbb{J}$ if it belongs to $int
K(\Bbb{J})$.

For stable dissipative operators the remark 2 is true: if an
operator is stable dissipative with respect to the family of
compacts and the interior of at least one of them is not empty,
then it is stable dissipative with respect to some norm.

\subsection {The Moore Effect for Autonomous Systems}

The results of the previous section can be applied to the study of
the Moore sweep effect. First give the exact definition of what we
understand by the Moore effect.

Let in the vicinity of a compact convex body $B \subset \Bbb{R}^n$
be given a smooth autonomous system $$ \frac {dx}{dt}=f(x) \eqno
{(7)} $$ with $B$ positively invariant with respect to (7), and
let $x(0)$ be determined inexactly, namely $$ x(0) \in x_0 + W_0,
$$ where $x_0 \in B, \  W_0 \in \Bbb{J}, \  x_0+W_0 \in B, \
\Bbb{J}$ is some interval space (see definition 1).

{\bf Remark 7}. Irrespective of particular numerical method (i.e.
dealing with the exact solution of the initial value problem for
(7) with the initial conditions \mbox{$x(0)=x_0$}) a stepwise
interval solution with step $h>0$ can be described as follows.

Let $T_h$ be the transformation of the phase flow of (7) during
the time $t$ ({\it shift over time} $t$), $W_0 \in \Bbb{J}$ is the
initial interval (its sense is an uncertainty in initial data).
Assume $$X_0=x_0+W_0,$$ $$X_{m+1}=T_{(m+1)h}x_0 + W_{m+1},$$
$$W_{m+1}=\bigcap_{W \supset W_{m+1}(h), W \in \Bbb{J}} W,$$
$$W_{m+1}(h)=T_h(T_{mh}x_0+W_m)-T_{(m+1)h} x_0 .$$

The sequence $\{ X_{m} \}_{m=0}^{+\infty}$ is the {\it exact
stepwise interval solution} of (7).

{\bf Definition 6}. {\bf The absence} of {\it infinitesimal Moore
effect} (IME) means that for any $h>0$ the sequence
${\{W_m\}}_{m=0}^{+\infty }$ is enclosed: $W_m \supset W_{m+1}$
for all $m$, i.e. the obtained intervals do not expand.

With IME the intervals expand along any trajectory (7) for any
small step, and that means that when solving a system by a
stepwise interval numerical method with any small step the
interval expansion takes place for any initial data irrespective
of the applied method (since it is true even for exact solutions).

Generalizing the construction [10] for norms, introduce the
following functional: $$ N_W (x, y)=\lim_{h \to +0} \frac {{\mu
}_W (x+hy)-{\mu }_W (x)}{h}. $$

Literally (with substitution of the norm for Minkovski functional)
repeating the reasonings from [10] (pp.127, 426), come to the
following statemets.

{\bf Statement 1}. If $x(t)$ with values in $\Bbb{R}^n$ is
differentiable on connected subset $T$ of the real axis, and $W$
is a convex compact $(0 \in ri W)$, then the function ${\mu }_W
(x(t))$ is almost everywhere differentiable on $T$ and the
derivative (where it exists) coincides with the right-hand
derivative, equal to $N_W (x(t), \dot s(t))$. The right-hand
derivative of ${\mu }_W (x(t))$ exists everywhere on $T$ except
the right-hand end.

{\bf Statement 2}. $$ {\gamma }_W (A)= \sup_{x \in W} N_W (x, AX).
$$

By $f'(x)$ further on we denote the mapping derivative of $f$.

The main part of further results on IME can be obtained from the
following theorem.

{\bf Theorem 1}. Let in the region $U \subset R^n$ be given a
smooth autonomous system (7), $B \subset U$ be positively
invariant with respect to (7) compact convex body. IME is absent
for compact $B$, system (7) and interval space $\Bbb{J}$ if and
only if $$ f'(x) \in K(\Bbb{J}) \eqno {(8)} $$ for all $x \in B$,
i.e. for any $x\in B$ the Jacoby matrix of system (7) in the point
$x$ is strongly dissipative with respect to $\Bbb{J}$.

{\bf Proof}. {\it Sufficiency}. Let $W \in \Bbb{J}$. Consider two
solutions $x_1 (t), x_2 (t)$ of system (7) with initial conditions
from B. Denote $\Delta (t)=x_1(t)-x_2(t)$. Using statements 1 and
2 and the theorem on finite increment, estimate the derivative of
${\mu }_W (\Delta (t))$: $$ \frac {d}{dt}{\mu }_W (\Delta (t))=
N_W (\Delta (t), d\Delta (t)/dt) \leq$$ $$ \leq \sup_{0 \leq
\Theta \leq 1} N_W(\Delta (t), f'(x_c(t))\Delta(t)) \leq $$ $$\leq
\sup_{0 \leq \Theta \leq 1}{\gamma }_W(f'(x_c(t))) {\mu }_W(\Delta
(t)),$$ where $x_c(t)=x_1(t)+\Theta (x_2(t)-x_1(t)), \ 0 \leq
\Theta \leq 1$ for all $t \geq 0$. By (8) and statement 1 we
obtain $$ \frac {d}{dt}{\mu }_W (\Delta (t)) \leq 0. $$

Since the latter inequality holds for all $t \geq 0$ and for all
$W \in \Bbb{J}$, in systen (7) on $B$ IME with respect to
$\Bbb{J}$ is absent.

{\it Necessity}. Let $W \in \Bbb{J},\  x_0 \in int B,\  t_0 \geq
0,\ y \in AffB,\  y \ne 0$. There exists such $h_0 >0$ that
$x_0+h_0y \in B$. Due to smoothness of system (7) there exist and
are unique the solutions $x_1(t), x_2(t)$ of the initial value
problem for (7) with the initial conditions $x_1(t_0)=x_0,\
x_2(t_0)=x_0+h_0y$. Assume $\Delta (t)=x_1(t)-x_2(t)$. Then $$
\frac {d}{dt} \ln {{\mu }_W} (\Delta (t)){\vert }_{t=t_0}= $$ $$ =
N_W \biggl( \frac {\Delta (t_0)}{{\mu }_W(\Delta (t_0))},
f'(x_c)\frac {\Delta (t_0)}{{\mu }_W(\Delta (t_0))}\biggr) = $$ $$
= N_W\biggl( \frac {y}{{\mu }_W(y)}, f'(x_c)\frac {y}{{\mu
}_W(y)}\biggr), $$ where $x_c=x_0+\Theta h_0y,\ \  0 < \Theta <1$.

By virtue of absence of IME $$ \frac {d}{dt}\ln {\mu }_W(\Delta
(t)) \leq 0 $$ for all $t \geq 0$. Since if $x_0+h_0y \in b$,
then:\\ (a) $x_0+hy \in B$ for all $h \in [0; h_0]$,\\ (b) a set
of those $h \in [0; h_0]$, for which $$ N_W \biggl( \frac {y}{{\mu
}_W(y)}, f'(x) \frac {y}{{\mu }_W(y)}\biggr) \leq 0, $$ is dense
on the segment $[0; h_0],$\\ and (c) due to its closureness
coincides with this segment.

By virtue of arbitrarity of the choice of $x_0$ for any $x_0 \in
int B,$ $t \geq 0,$ \mbox{$y \in AffB$}, \linebreak[1] \mbox{$y
\ne 0$} the inequality $$ N_W \biggl( \frac {y}{{\mu }_W(y)},
f'(x)\frac {y}{{\mu }_W (y)}\biggr) \leq 0 $$ is satisfied. It
holds also for any $x \in B, t \geq 0, y \in AffB, y \ne 0$.
Hence, from lemma 1 and statement 2 immediately follows
dissipativity of $f'(x)$ with respect to $\Bbb{J}$ for all $x \in
B$. The theorem is proved.

{\bf Definition 7}. The family of linear operators $\{A_\alpha \}$ is called {\it
simultaneously dissipative}, if there exists a norm with respect to which all the
operators are dissipative.

Simultaneously dissipative operators were studied in detail in
[11, 12, 17-22].

From theorem 1, example 3, and remark 2 we obtain the following
theorem.

{\bf Theorem 2}. For existence of interval space in which at least
one interval posesses non-empty interior and with respect to which
in system (7) there is no IME on $B$, it is necessary and
sufficient for the family $\{f'(x): x \in B \}$ to be
simultaneously dissipative.

Thus, the problem of existence of the interval space, with respect to which IME is
absent, is reduced to the problem of simultaneous dissipativity of Jacobi matrices. As
sought for space one can choose a set of all balls of that norm with respect to to which
all Jacobi matrices are dissipative. This norm is {\it contracting} for (7) on $B$ (i.e.
the distance between two solutions with initial conditions from B will not expand with
time). Hence, all systems without IME (with respect to some interval space) on $B$ are
globally stable in $B$ (see introduction).

Bellow by $C^1(B)$ we denote the Banach space of smooth mappings
of $B$ in $\Bbb{R}^n$ with the norm $$ {\Vert f \Vert}_{C^1(b)} =
\max_{x \in B} \Vert f(x) \Vert + \sum_{k=1}^{n} \max_{x \in B}
\biggl\| \frac{\partial f}{\partial x_k} \biggr\| $$ where $\Vert
. \Vert$ is a fixed norm in $\Bbb{R}^n$.

Further on, speaking about properties of autonomous systems, we
mean the properties of the vector fields generating them.

Immediately from lemma 3 and theorem 1 the following statement can
be obtained.

{\bf Theorem 3}. The set of systems on $B$ without IME with
respect to $\Bbb{J}$ is closed convex cone in $C^1(B)$.

For this cone we use the notation $F_B(\Bbb{J})$.

Further on, speaking about the vicinity of an autonomous system in
$C^1(B)$ we mean a part of the vicinity, consisting only of those
systems for which the set $B$ is positively invariant.

Let us study under what conditions the interior of the cone
$F_B(\Bbb{J})$ is non-empty.

{\bf Theorem 4}. For non-emptiness of $int F_B(\Bbb{J})$ in
$C^1(B)$ it is necessary and sufficient for all elements of
$\Bbb{J}$, exept $\{0\}$, to possess non-empty interior.

{\bf Proof}. {\it Necessity}. Let exist such a set $W \in \Bbb{J}$
that $int W=0$. Consider any system (7) without IME with respect
to $\Bbb{J}$ on $B$. Since $int B \ne 0$, there exist two
different concentrical balls $S_1$ and $S_2$ of usual $l^2$-norm,
belonging to $int B$ with $S_1 \subset S_2$. Construct such a
function $g \in C_{\infty} (\Bbb{R}^n)$ that $g(x)=1$ for all $x
\in S_1$ and $g(x)=0$ at all $x \notin S_2$. Since $AffW \ne
\Bbb{R}^n$, one can construct a linear operator $A \in
L(\Bbb{R}^n)$ mapping $AffW$ into such a subspace $E_0 \ne \{0\}$
that $(AffW) \cap E_0 = \{0\}$.

Consider the system $$ \frac {d x}{d y} = f(x)+\varepsilon g(x)AX, \eqno {(9)} $$ where
$\varepsilon > 0$ is arbitrary. The set $B$ is positively invariant with respect to (9),
since the vector field generating (9) coincides with $f$ in the vicinity of $\partial B$.
On the other hand, there exist Jacobi matrices (9) with respect to which $AffW$ is not
invariant, i.e. in (9) exist IME with respect to $\Bbb{J}$ on $B$. Since in any vicinity
of $f$ there is at least one vector field, generating (9), then $$ int F_B(\Bbb{J})=0. $$

{\it Sufficiency}. Consider the system $dx/dt=-x$. It is a system
on $B$ without IME with respect to $\Bbb{J}$. Furthermore, if all
elements of $\Bbb{J}$, except $\{0\}$, possess non-empty interior,
then by lemma 5 the matrix of the system is stable dissipative
with respect to $\Bbb{J}$ (see definition 5).

Consider the system: $$ \frac {d x}{d y} = -x+v(x), \eqno {(10)}
$$ where ${\Vert v \Vert }_{C_1(B)} < \varepsilon $ with
$\varepsilon $ chosen so that $$ A - I \in K(\Bbb{J}) $$ if $\Vert
A \Vert < \varepsilon$ (see the proof of lemma 5). Then all Jacobi
matrices (10) are dissipative with respect to $\Bbb{J}$, and if
$v$ is choosen so that $B$ is positively invariant with respect to
(10), then in (10) IME is absent (by theorem 1). The theorem is
proved.

Thus, $int F_B(\Bbb{J}) \ne 0$ if and only if $int K(\Bbb{J}) \ne
0$.

It is easy to see that in the proof of sufficiency in theorem 4
one can instead of the system $dx/dy=-x$ consider any system whose
Jacobi matrices are stable dissipative with respect to $\Bbb{J}$.

{\bf Remark 8}. By analogy with the space $C^1(B)$ one can
construct the Banach spaces $C^k(B) \  (k \in N)$ with the norm $$
{\Vert f \Vert }_{C^k(B)}=\sum_{\vert \alpha \vert =0}^{k} \max_{x
\in B}{\Vert (D^{\alpha } f)(x) \Vert }, $$ where $\alpha =(\alpha
_1, \ldots , \alpha _n)$ is a multiindex: $$ \vert \alpha \vert =
\alpha _1 + \ldots + \alpha _n, D^{\alpha }f= \frac {{\partial
}^{\vert \alpha \vert } f} {\partial x_1^{\alpha _1} \ldots
\partial x_n^{\alpha _n}} $$ and the metric space $C^{\infty }(B)$
with the system of seminorms $$ \{\max_{x \in B} \| (D_\alpha
f)(x)\| : \  \vert \alpha \vert \leq m \}_{m=0}^{+\infty }. $$
Small $C^k$-additions $(1 \leq k \leq +\infty )$ are small and in
the $C^1$-norm. Therefore, for $C^k$-smooth systems under the
conditions of theorem 4 the interior of $F_B(\Bbb{J})$ is
non-empty and in $C^k(B)$. As shows the proof of theorem 4
(necessity), if conditions of the theorem are not satisfied,then
the interior of $F_B(\Bbb{J})$ in $C^k(B)$ is empty.

Let clarify what autonomous system without IME in specific
interval spaces looks like.

{\bf Theorem 5}. Any system without IME with respect to $\Bbb{J}$
from example 1 has the form: $$ \frac {d x}{d t} = ax+c, $$ where
$a \leq 0,\  C \in \Bbb{R}^n$ is a constant vector.

{\bf Proof}. Let $A \in K(\Bbb{J})$. All the segments symmetrical
with respect to $0$ belong to $\Bbb{J}.$ Every such a segment has
the form $\{y \in \Bbb{R}^n | y=ax, \vert a \vert \leq 1 \}$ for
some $x \in \Bbb{R}^n.$ The cone $Q_x$ (see lemma 2) for each
segment consists of vectors of the form $ax$, where $a<0$. Thus,
every non-zero vector $x \in \Bbb{R}^n$ is eigenvector of the
operator $A$, corresponding to non-positive eigenvalue. Thus: $$
K(\Bbb{J})=\{aI \vert \  a \leq 0\}. \eqno {(11)} $$

Let now a system without IME have the form\\ $$ \left \{
\begin{array}{rcl}
\displaystyle {dx_1 \over dt} & = & f_1(x_1, \ldots , x_n); \\
 &\cdots &\\
\displaystyle {dx_n \over dt} & = & f_n(x_1, \ldots , x_n) .
\end{array}
\right. $$

According to (11) and theorem 1 $$ \frac {\partial f_i}{\partial
x_j} \equiv 0 \  (i \ne j); \eqno{(12)} $$ $$ \frac {\partial
f_1}{\partial x_1}\equiv \frac {\partial f_2}{\partial x_2}\equiv
\ldots \equiv \frac {\partial f_n}{\partial x_n} \leq 0. \eqno
{(13)} $$

From (12) follows that $f_k$ depends only on $x_k \  (k=1, \ldots
, n)$. It means that ${\partial f_k}/{\partial x_k}$ also depends
only on $x_k$, i.e. by virtue of (13) ${\partial f_k}/{\partial
x_k}=const \  (k=1, \ldots , n)$. Then $$ \frac {\partial
f_1}{\partial x_1}\equiv \frac {\partial f_2}{\partial x_2}\equiv
\ldots \equiv \frac {\partial f_n}{\partial x_n} \equiv a \leq 0
$$ and the system has the form: $$ \frac {d x}{d t} = ax + c, $$
where $a \leq 0, \ c=const$. The theorem is proved.

Thus, whatever nonlinear (or even linear with non-scalar matrix)
system we consider, if we take as $\Bbb{J}$ the interval space of
example 1 (or any wider space), IME will be present in the system.
From theorem 5 also follows that any dissipative with respect to
all norms operator has the form $aI$, where $a \leq 0$ (see also
remark 3).

{\bf Example 6}. Consider $\Bbb{J}$ from example 2. $\Bbb{J}$
contains all symmetrical with respect to $0$ segments of
coordinate axes (thus, the conditions of theorem 4 are not
satisfied, i.e. $int F_B(\Bbb{J})=0$). Let $A \in K(\Bbb{J})$.
Reasoning like in proof of theorem 5, conclude that all coordinate
axes are eigenspaces of the operator $A$, corresponding to
non-positive eigenvalues. In other words, the matrix of the
operator $A$ is diagonal and non-positive. On the other hand, by
virtue of lemma 2, all such operators belong to $K(\Bbb{J})$.
Thus, systems without IME with respect to $\Bbb{J}$ on $B$ have
the form $$ \left \{
\begin{array}{ccc}
\displaystyle {dx_1 \over dt} & = & f_1(x_1); \\ [4mm]
 &\cdots &\\
\displaystyle {dx_n \over dt} & = & f_n(x_n)
\end{array}
\right. $$ where $$ \frac {\partial f_k}{\partial x_k} \leq 0 \
(k=1, \ldots , n) $$ for all $x \in B$.

From the considered example follows that when using standard
intervals (rectangular parallelepipeds) IME will be observed in
almost all systems in $\Bbb{R}^n \ \mbox{if} \ n \neq 1.$

The systems without IME with respect to $\Bbb{J}$ from example 3 on $B$ represent all
systems for which the norm $\Vert . \Vert $ is contracting in $B$ (see the text after
theorem 2).

{\bf Remark 9}. Note that testing of dissipativity (stable
dissipativity) of the operator with respect to the norm is
equivalent to non-positiveness (negativeness) of the corresponding
Lozinsky norm. For some norms an explicit form of corresponding
Lozinsky norm is known (see, for example, [9] or [10, p.463-465]).
In particular, for the Euclidean norm the Lozinsky norm of the
operator $A$ coincides with the largest eigenvalue of the operator
$(A^*+A)/2$. The Lozinsky norm of the operator $A$ represented by
the matrix $(a_{ij})_{i,j=1}^n$ with respect to $l^1-$ and
$l^\infty $-norms is given by the formulae, respectively: $$
\max_{1 \leq i \leq n} (\mbox{Re}\ a_{ii}+\sum_{j \ne i} \vert
a_{ji} \vert); $$ $$ \max_{1 \leq i \leq n} (\mbox{Re}\
a_{ii}+\sum_{j \ne i} \vert a_{ij} \vert). $$

In remark 9 it is assumed that the operator $A$ acts in the space
$C^n$. The definitions and used here properties of dissipative
operators in complex spaces are analogous to those in real ones.

{\bf Example 7}. Let $\Bbb{J}$ be the interval space from example
4. Then $$ K(\Bbb{J})=\bigcap_{1 \leq k \leq m} K_{{\Vert .
\Vert}_k}, $$ where $K_{{\Vert . \Vert}_k}$ is the cone of all
operators dissipative with respect to the norm $K_{{\Vert .
\Vert}_k}$. It follows from the closure under intersection of the
family of all positively invariant sets of an autonomous system.
Similarly, $$ int K(\Bbb{J})=\bigcap_{1 \leq k \leq m} int
K_{{\Vert . \Vert}_k}. $$

Let, for example, $\Vert . \Vert _1$ be $l^\infty $-norm, $\Vert .
\Vert _2$ be $l^2$-norm in $\Bbb{R}^2$. Then the conditions of
stable dissipativity of the operator $A$ with the matrix
$(a_{ij})_{i,j=1}^2$ with respect to $\Bbb{J}$ according to remark
9 are of the form: $$ \left \{
\begin{array}{lcl}
4a_{11}a_{22} & > & (a_{12}+a_{21})^2; \\ [4mm] a_{11}+\vert
a_{12} \vert & < & 0; \\ [4mm] \vert a_{21} \vert +a_{22} & < & 0.
\end{array}
\right. $$

Thus, for the system of the form $$ \left \{
\begin{array}{rcl}
\displaystyle {dx_1 \over dt} & = & f_1(x_1, x_2); \\ [4mm]
\displaystyle {dx_2 \over dt} & = & f_2(x_1, x_2)
\end{array}
\right. \eqno {(14)} $$ if the inequalities $$ \left \{
\begin{array}{lcl}
4\displaystyle {{\partial f_1} \over {\partial x_1}}\cdot
 \displaystyle {{\partial f_2} \over {\partial x_2}} & > &
\left( \displaystyle {{\partial f_1} \over {\partial x_2}}+
\displaystyle {{\partial f_2} \over {\partial x_1}}\right)^2 ;\\
[4mm] \displaystyle {{\partial f_1} \over {\partial x_1}}+\biggl|
\displaystyle {{\partial f_1} \over {\partial x_2}} \biggr| & < &
0;\\ [4mm] \biggl| \displaystyle {{\partial f_2} \over {\partial
x_1}} \biggr| + \displaystyle {{\partial f_2} \over {\partial
x_2}} & < & 0
\end{array}
\right. $$ are satisfied and the compact convex body $B$ is
positively invariant with respect to (14), then in (14) IME with
respect to $\Bbb{J}$ (from example 4) is absent on $B$. For
example, such is the following system:

$$ \left \{
\begin {array}{rcl}
\displaystyle{dx_1 \over dt} & = & -2x_1+x_2;\\ [4mm]
\displaystyle{dx_2\over dt} & = & 2x_1-3x_2
\end{array}
\right. $$ if $B$ is the square $\{(x_1, x_2):\  \vert x_1 \vert
\leq 1, \vert x_2 \vert \leq 1 \}$ or the circle $\{(x_1, x_2): \
x_1^2 = x_2^2 \leq 1\}$.

{\bf Example 8}. Consider $\Bbb{J}$ from example 5. Let $Q$ be
rectangular triangle with vertices at the points $(-1;2); (-1;-1);
(1;-1)$.

From lemma 2 and theorem 1 follows that the cone $F_B(\Bbb{J})$
consists of the systems of the form (14), with respect to which
the compact $B$ is positively invariant and for which $$ \left \{
\begin {array}{lcl}
\displaystyle{{\partial f_1} \over {\partial x_1}} +
\displaystyle{{\partial f_1} \over {\partial x_2}} & \leq & 0;\\
[4mm] \displaystyle{{\partial f_1} \over {\partial x_1}} -2
\displaystyle{{\partial f_1} \over {\partial x_2}} & \leq & 0;\\
[4mm] \displaystyle{{\partial f_2} \over {\partial x_1}} + \biggl|
\displaystyle{{\partial f_2} \over {\partial x_2}} \biggr| & \leq
& 0; \\ [4mm] -3\displaystyle{{\partial f_1} \over {\partial x_1}}
+
 6\displaystyle{{\partial f_1} \over {\partial x_2}} -
 2\displaystyle{{\partial f_2} \over {\partial x_1}} +
 4\displaystyle{{\partial f_2} \over {\partial x_2}} & \leq & 0; \\ [4mm]
 3\displaystyle{{\partial f_1} \over {\partial x_1}} -
 3\displaystyle{{\partial f_1} \over {\partial x_2}} +
 2\displaystyle{{\partial f_2} \over {\partial x_1}} -
 2\displaystyle{{\partial f_2} \over {\partial x_2}} & \leq & 0.
\end{array}
\right.\eqno(15) $$ is true.

Substituting all the inequality signs in (15) by strict ones,
obtain $int F_B(\Bbb{J})$. For example, the system $$ \left \{
\begin{array}{rcl}
\displaystyle{dx_1 \over dt} & = & -x_1; \\ [4mm]
\displaystyle{dx_2 \over dt} & = & -6x_1 - 4x_2
\end{array}
\right. $$ belongs to $int F_B(\Bbb{J})$ for $B=Q$.

Corresponding ball $S$ (see remark 2) is the parallelogram with
the vertices in the points $(-1;2); (-1;-1); (1;-2); (1;1)$. From
remark 2 follows that $$ K(\Bbb{J})=K(Q) \subset K(S). $$ One can
see that $K(\Bbb{J}) \ne K(S)$. For example, the operator given by
the matrix $$ \left(
\begin{array}{rr}
-4 & -1 \\ [4mm]
 2 & 0
\end{array}
\right) $$ is dissipative with respect to $S$, but it is not dissipative with respect to
$\Bbb{J}$. In other words, in the systems without IME with respect to $\{\alpha S:\
\alpha \geq 0\}$ (i.e. contracting according to the norm whose ball is $\Bbb{J}$) there
can be observed IME with respect to $\Bbb{J}$.

This example can be generalized as follows. Consider $\Bbb{J}$
from remark 2. In system (7) on $B$ IME is absent with respect to
$\Bbb{J}$ if the operators $f'(x)$ for all $x \in B$ are
dissipative with respect to all sets $Q_k (k=1, \ldots, m)$.

To sum up, one can say the following. When using sufficiently wide
interval spaces in almost all systems in accordance with theorem
4, IME is observed. In particular, IME takes place almost for all
systems when using standard intervals (see example 6). Expansion
of the interval space results in the appearance of new systems
with IME: thus, in using a set of all symmetrical to $0$ convex
compacts IME is absent only for linear systems with non-positive
scalar matrices. And the most impotent: the question about the
existence of interval space, with respect to which in the
considered system IME is absent, is reduced to the problem of
joint dissipativity of the Jacobi matrices. Therefore, there is no
interval space with respect to which all (or even if in some sense
almost all) globally stable systems would have no IME. One has to
solve individually problems of the existence and constructing of
corresponding interval spaces for each particular system. These
problems are solved constructively very rarely.

We have treated the Moore effect in a very strong sense. The condition of boundedness of
the sequence of intervals ${\{W_m\}}_{m=0}^{+\infty }$ at any step $h>0$ (see remark 7)
is weaker (and acceptable, generally speaking, for constructing sufficiently narrow
interval solutions). This condition can be called the condition of absence of the {\it
asymptotic Moore effect} (AME). It is the weakest from acceptable conditions, since with
AME it is impossible to use stepwise interval methods to obtain narrow interval solutions
at large times. The study of AME is still not completed. It is evident only that for a
linear autonomous system in considering the interval space from example 3 AME is
equivalent to IME. One can suggest a hypotesis: the problem of existence and constructing
of the interval space with respect to which AME is absent in the autonomous system is
reduced to the question of simultaneous dissipativity of Jacobi matrices (and of
constructing a contracting norm).

\section{Conditions of Simultaneous Dissipativity of Operators}
\subsection{Some General Results}

In the present section some conditions of simultaneous
dissipativity of the operators will be considered (see definition
7).

A definition of a simultaneous dissipativity can be generalized in
such a way.

{\bf Definition 7$'$}. A family of linear operators $\{A_\alpha
\}$ is called {\it simultaneously stable dissipative} if there
exists a norm with respect to which all operators $A_\alpha $ are
stable dissipative.

{\bf Lemma 6}. Let the space $\Bbb{R}^n$ be expanded into direct
sum of subspaces $E_i \  (i=1, \ldots , k)$ and each of them is
invariant with respect to all operators of the family $\{A_\alpha
\}$. Further on, let restriction of the family  $\{A_\alpha \}$ on
any $E_i$ be simultaneously (sumultaneously stable) dissipative.
Then $\{A_\alpha \}$ is simultaneously (simultaneously stable)
dissipative.

{\bf Proof}. Let ${\Vert . \Vert }_i \  (i=1, \ldots, k)$ be the
norms in $E_i$ in which the restrictions of $\{A_\alpha \}$ on
$E_i$ are simultaneously (simultaneously stable) dissipative.
Define the norm in $\Bbb{R}^n$ in this way: $$ \Vert x \Vert =
\sum_{i=1}^k {\Vert x_i \Vert }_i, $$ where $x=\sum_{i=1}^k x_i$
with $x_i \in E_i \  (i=1, \ldots, k)$.

In this norm all operators $A_\alpha $ are simultaneously
(simultaneously stable) dissipative. The lemma is proved.

It is known [7] that for one operator the norm with respect to
which it is dissipative exists if and only if the spectrum of the
operator lies in the closed left half-plane and the boundary part
is diagonalizable (i.e. Jordan boxes corresponding to pure
imaginary, including zero ones, eigenvalues are diagonal). The
norm, with respect to which the operator is stable dissipative,
exists if and only if the spectrum of the operator lies in the
open left half-plane.

Several stable dissipative (in their own norms) operators not
necessarily are simultaneously dissipative. To demonstrate that,
consider operators represented by the matrices $$ A_1= \left(
\begin{array}{rr}
-1 & 3 \\ [4mm]
 0 & -1
\end{array}
\right); \ \ A_2= \left(
\begin{array}{rr}
-1 & 0 \\ [4mm]
 3 & -1
\end{array}
\right). $$ Each of them is stable dissipative in its norm (due to
the location of the spectrum). But $$ A_1+A_2= \left(
\begin{array}{rr}
-2 & 3 \\[4mm]
 3 & -2
\end{array}
\right). $$ The spectrum of the operator $(A_1 + A_2)$ contains
the point $\lambda =1$ which does not belong to the closed left
half-plane. Thus, the operator $(A_1 + A_2)$ is not dissipative in
any norm. By lemma 6 the operators $A_1$ and $A_2$ are not
simultaneously dissipative.

The problem to find out necessary and sufficient conditions of simultaneous dissipativity
of an arbitrary (even finite) family of operators seems to be very difficult.
Nevertheless, one can obtain some sufficient conditions imposing different constraints on
the operators. We obtain the sufficient condition of simultaneous dissipativity of the
family generating a solvable Lee algebra. let us recall [13] that a family of matrices
generates a solvable Lee algebra if and only if all elements of this family are
simultaneously reducible to triangular form (generally speaking in complex basis).

{\bf Theorem 6}. Let the family $\{A_\alpha \}$ be compact and
generate solvable Lee algebra, and the spectrum of each operator
$A_\alpha $ lies in the open left half-plane. Then $\{A_\alpha \}$
is simultaneously stable dissipative.

{\bf Proof}. First consider the case of complex space $\Bbb{C}^n$.
Consider matrices of the operators $A_\alpha $ in the basis where
they are of triangular form.

Let each matrix $A_\alpha $ have the form $$ A_\alpha = \left(
\begin{array}{cccccc}
\lambda_1^{(\alpha )} & 0 & 0 & \cdots & 0 & 0 \\ [4mm]
\mu_{21}^{(\alpha )} & \lambda_2^{(\alpha )} & 0 & \cdots & 0 & 0
\\ [5mm] . & . & . & \cdots & 0 & 0 \\ [5mm] \mu_{n1}^{(\alpha )}
& \mu_{n2}^{(\alpha )} & \mu_{n3}{(\alpha )} & \cdots &
\mu_{n,(n-1)}^{(\alpha )} & \lambda_n^{(\alpha )}
\end{array}
\right). $$

Show the existence of such a set of positive numbers
$\{c_k\}_{k=1}^n$ that all $A_\alpha $ are stable dissipative in
the norm $$ \Vert z \Vert =\max_{1\leq k\leq n}\frac {\vert z_k
\vert}{c_k} \eqno{(16)} $$ (here $z_k$ is the $k$-th coordinate of
the vector $z$ in the given basis), whose unit ball is the
polycylinder $$ \vert z_k \vert \leq c_k \  (k=1, \ldots , n).
\eqno {(17)} $$

If $\{e_k\}_{k=1}^n$ is the considered basis, then, evidently,
norm (16) coincide with the $l^{\infty }$-norm with respect to the
basis $\{c_k / e_k\}_{i,j=1}^n$ in the norm (16): $$ \mbox{Re}\
a_{ii} + \sum_{j \ne i}\frac {c_j}{c_i}\vert a_{ij} \vert < 0 \
(i=1, \ldots, n). \eqno {(18)} $$

For the matrices $A_\alpha $ the conditions (18) look like this:
$$ \left \{
\begin {array}{lcl}
\mbox{Re}\ \lambda_1^{(\alpha )} & < & 0 ; \\ [4mm] \mbox{Re}\
\lambda_2^{(\alpha )} + \displaystyle {c_1 \over c_2} \vert
\mu_{21}^{(\alpha )}\vert & < & 0 ; \\ [4mm] & \cdots & \\ [4mm]
\mbox{Re}\ \lambda_n^{(\alpha )} + \displaystyle {c_1 \over c_n}
\vert \mu_{n1}^{(\alpha )}\vert + \ldots + \displaystyle {c_{n-1}
\over c_n} \vert \mu_{n, (n-1)}^{(\alpha )} \vert & < & 0 . \\
[4mm]
\end{array}
\right. \eqno {(19)} $$

Suppose $\mu =\sup_{\alpha,k\ne l} \vert \mu_{kl}^{(\alpha
)}\vert; \  \lambda =-\sup_{\alpha,k} \mbox{Re}\
\lambda_k^{(\alpha )}$. From the conditions of the theorem follows
that $0 < \lambda < +\infty , \  0 < \mu < +\infty $. To fulfil
(19) for all $A_\alpha $, it is sufficient that the inequalities
$$ (c_1 + \ldots + c_{k-1}) \mu < c_k \lambda \ (k=1, \ldots , n);
\  c_1 > 0 \eqno {(20)} $$ be satisfied.

Show the solvability of system (20). Let $c_1=1$. Choose the
others $c_k$ so that $$ c_2 > \mu/\lambda; c_3 >
(1+c_2)\mu/\lambda; \ldots;$$ $$ c_n > (1+c_2+\ldots
+c_{n-1})\mu/\lambda.$$ Then the inequalities (20) are satisfied,
i.e. all operators $A_\alpha $ are stable dissipative in the norm
(14).

Let now operators $A_\alpha $ act in the space $\Bbb{R}^n$. In
usual way complexify $\Bbb{R}^n$ and the family ${A_\alpha}$.
Then, as it has been described above, construct a cylinder (17).
Intersection of (17) with the initial space $\Bbb{R}^n$ produce a
ball of the norm in which all $A_\alpha $ are stable dissipative.
The theorem is proved.

If instead of stable dissipative operators one considers
dissipative operators, then the analog of theorem 6 is not true,
starting from real dimension 4. Let $$ A_1= \left(
\begin{array}{rr}
i & 1 \\[4mm] 0 & 2i
\end{array}
\right); A_2= \left(
\begin{array}{rr}
2i & 1 \\[4mm] 0 & i
\end{array}
\right). $$ Each of the operators $A_{1,2}$ is dissipative in its
norm. The finite family is compact, the matrices $A_1$ and $A_2$
generate solvable Lee algebra. Nevertheless $$ A_1 + A_2 = \left(
\begin{array}{rr}
3i & 2 \\[4mm] 0 & 3i
\end{array}
\right). $$

The only eigenvalue of the operator $(A_1+A_2)$ is pure imaginary,
with the matrix of this operator representing (up to a constant
factor) non-trivial Jordan box. That means it is not dissipative
in any norm, i.e. $A_1$ and $A_2$ are not simultaneously
dissipative. To obtain a real example, one has to make the
matrices $A_1$ and $A_2$ real: $$ A_1^R= \left(
\begin{array}{rrrr}
0 & -1 & 1 & 0 \\ [4mm] 1 & 0 & 0 & 1 \\ [4mm] 0 & 0 & 0 &-2 \\
[4mm] 0 & 0 & 2 & 0
\end{array}
\right); A_2^R= \left(
\begin{array}{rrrr}
0 & -2 & 1 & 0 \\[4mm] 2 & 0 & 0 & 1 \\[4mm] 0 & 0 & 0 &-1 \\[4mm]
0 & 0 & 1 & 0
\end{array}
\right). $$

To keep true the statement about simultaneous dissipativity for
nonstable dissipative operators, it is sufficient to strengthen
the requirement of solvability up to nilpotency. Remind [13] that
for each linear operator $A$ in the space $E$ the operator $ad\ A$
in $L(E)$ is defined: $$ (ad\ A)B=AB-BA. $$

The family $\{A_\alpha \}$ generates the nilpotent Lee algebra if
and only if there exists such a number $m\in N$ that for any set
of $\{A_{\alpha _k}\}_{k=1}^m$ (among the elements of wich there
may be the same ones) and for all $\alpha $: $$ \prod_{k=1}^m (ad\
A_{\alpha_k}) A_\alpha = 0. \eqno {(21)} $$

Nilpotent Lee algebra is always solvable. Commutative Lee algebra
is nilpotent (for it $m=1$) and solvable.

{\bf Theorem 7}. Let the family $\{A_k\}$ be finite and generate
nilpotent Lee algebra, and for each operator $A_k$ exist a norm
with respect to which it is dissipative. Then $\{A_k\}$ is
simultaneously dissipative.

{\bf Proof}. Without loss of generality one can assume that among
the operators $A_k$ there are no scalar ones (if $A=aI,$ where
$\mbox{Re}\ a \leq 0,$ then $A$ is dissipative in any norm) and
exists at least one operator (denote it $A_1$), among eigenvalues
of which there are pure imaginary (otherwise we are under the
conditions of theorem 6).

First assume that $A_k$ operates in $\Bbb{C}^n$. We prove the
theorem by induction on dimension of space. In dimension 1 the
statement of the theorem is trivial. Show that one can expand all
the space $\Bbb{C}^n$ into a direct sum of two non-trivial
subspaces invariant with respect to all $A_k$. Since in both of
them the conditions of the theorem (for corresponding restrictions
of $\{A_k\}$) are satisfied, then to complete the proof one has to
use lemma 6.

Let $\lambda $ be an imaginary eigenvalue of $A_1$; $E'$ be the
corresponding to $\lambda $ eigen-subspace (by virtue of
diagonalizability of boundary part of $A_1$ it coincides with
whole corresponding root subspace); $E''$ be the sum of root
subspaces corresponding to all the others eigenvalues of $A_1$.
Evidently, $\Bbb{C}^n=E'\oplus E''$ (the sign $\oplus $ means
direct sum); $E' \ne \Bbb{C}^n$, otherwise the operator $A_1$ is
scalar. Show the invariance of $E'$ and $E''$ with respect to all
$A_k$.

Let $x \in E'$. Then $$ A_1x=\lambda x. $$

On the other hand, in accordance with (21) there exists such $m
\in N$ that $(ad\ A_1)^m A_k=0$ for all $k$ and $$ (A_1 - \lambda
I)^m A_k x = 0. $$

A more general fact is true: if $Ax=0$ and $(ad\ A)^mB=0,$ then
$A^mBx=0$. For $m=0$ the fact is obvious. Let that be true for
$m=r$. Assume $$ Ax=0; (ad\ A)^{r+1}B=0. $$

Then $(ad\ A)^r (ad\ A)B=0$, and according to the inductive
hypothesis \linebreak[4] $A^r(ad\ A)Bx=0.$ But
$A^{r+1}Bx=A^r(BAx+(ad\ A)Bx)$, i.e. $A^{r+1}Bx=0$, as was to be
proved.

As a consequence of coincidence of $E'$ with the whole root
subspace, corresponding to $\lambda,$ we have: $$ (A_1-\lambda
I)A_kx=0, $$ i.e. $A_1A_kx=\lambda A_kx, A_kx \in E'$.

Show now the invariance of $E''$. Let $\{e_j\}_{j=1}^n$ be the
Jordan basis of the operator $A_1$ with $E'$ being corresponded to
the vectors $\{e_j\}_{j=j_1}^{j_2}$. One has to show that for any
$j$ less then $j_1$ or more then $j_2$ the coordinates of $A_ke_j$
with the numbers from $j_1$ to $j_2$ with respect to the assigned
basis are equal to zero. Let it be not so and exist such $j'$ that
$e_{j'} \in E''$, but the $j_1$-th coordinate $(j_1 \leq j_0 \leq
j_2)$ of the vector $A_ke_{j'}$ is $a \ne 0$. Write it like this:
$$ A_ke_{j'}= \ldots = ae_{j_0}. $$

Let $e_{j'}$ be an eigenvector of $A_1$ corresponding to the
eigenvalue $\mu \ne \lambda $. Then $$ (ad\
A_1)A_ke_{j'}=A_1(\ldots +ae_{j_0})-\mu (\ldots +ae_{j_0})= \ldots
+(\lambda -\mu )ae_{j_0}. $$

Verify that $$ (ad\ A_1)^m A_ke_{j'}= \ldots +(\lambda - \mu
)^mae_{j_0}. $$ For $m=0$ it is obvious. Let it be satisfied for
$m=r$. Then $$ (ad\ A_1)^{r+1}A_ke_{j'}=(ad\ A_1)(ad\
A_1)^rA_ke_{j'}= $$ $$ =A_1(ad\ A_1)^rA_ke_{j'}-(ad\
A_1)^rA_kA_1e_{j'}= $$ $$ =A_1(\ldots +(\lambda -\mu
)^rae_{j_0})-\mu (ad\ A_1)^rA_ke_{j'}= $$ $$ =\ldots +(\lambda
-\mu )^ra\lambda e_{j_0}-\mu (\lambda -\mu)^rae_{j_0}= $$ $$ =
\ldots +(\lambda -\mu )^{r+1}ae_{j_0}, $$ i.e. that is true also
for $m=r+1$, and, hence, for all $m \in N$.

Thus, $$ (ad\ A_1)^mA_ke_{j'}=\ldots +(\lambda -\mu )^mae_{j_0}
\ne 0 $$ for any $m \in N$, which contradicts (21).

Let now $e_{j'}$ be a root (but not eigen) vector, corresponding
to the eigenvalue $\mu $, with the $j_0$-th coordinate of the
vector $A_ke_{j'-1}$ equal to $0$. Then $$ (ad\
A_1)A_ke_{j'}=A_1(\ldots +ae_{j_0})- A_k(e_{j'-1} + \mu
e_{j'})=\ldots +(\lambda -\mu )ae_{j_0}. $$

Analogously $$ (ad\ A_1)^mA_ka_{j'}=\ldots +(\lambda -\mu
)^mae_{j_0} \ne 0 $$ for any $m \in N$, which contradicts (21).

Since the sequence of basis vectors belonging to the root subspace
begins with the eigenvector, the required statement for complex
space is proved.

The transfer onto the case of real space can be done in the same
way as in the proof of theorem 6 (the ball of corresponding norm
in the copmlexified $\Bbb{R}^n$ intersects with $\Bbb{R}^n$). The
theorem is proved.

From theorems 6 and 7 follows, in particular, that a finite
(compact) commutative family consisting of operators dissipative
(stable dissipative) in their own norms is simultaneously
dissipative (simultaneously stable dissipative).

\subsection{The Mass Action Law and Dissipative Mechanisms}

Some constructive conditions of simultaneous dissipativity can be
obtained for finite families of operators of rank 1. The problem of
the absence of Moore effect in the system constructed in
accordance with the {\it Mass Action Law} (MAL) is reduced to the
problem on simultaneous dissipativity of such operators.

MAL systems
appear from mathematical description of systems of chemical and
biological kinetics and in some other problems. To the considered
process is assigned an algebraic object, called {\it reaction mechanism}
and having the form:
$$
\alpha_{r1}A_1+\ldots +\alpha_{rn}A_n \to \beta_{r1}A_1+\ldots +
\beta_{rn}A_n,\  (r=1,\ldots , d). \eqno {(22)}
$$

Speaking in terms of chemical kinetics, the reaction
mechanism is a list of stoichiometric equations of
elementary reactions (22). In this case $A_1,\ldots , A_n$ are the
substances taking part in the reaction; $\alpha_{ri}, \beta_{ri}$
are the non-negative integers called stoichiometric coefficients and
showing in what amount the particles of $A_i$ enter into the $r$-th
elementary reaction as the initial substance $(\alpha_{ri})$ or
product $(\beta_{ri})$. The following notations are accepted:
$\gamma_{ri}=\beta_{ri}-\alpha_{ri},\  \gamma_r$ is the vector with
the components $\gamma_{ri} \  (i=1, \ldots , n)$ -- so-called
stoichiometric vector of the $r$-th elementary reaction.

In accordance with MAL [14,15], to the mechanism (22) corresponds
the following system of ordinary differential equations:
$$
\frac {dc_i}{dt}=\sum_{r=1}^d \gamma_{ri} w_r \eqno {(23)}
$$
where $c_i(t)$ is the concentration of substance $A_i$ at the moment of
time $t \geq 0,$
$$ w_r=k_r(t)\prod_{j=1}^n c_j^{\alpha_{rj}}$$
is the rate of the $r$-th elementary reaction, continuously
depending on time. In particular, if reaction proceeds under
constant external conditions, then $k_r(t)={\rm const}\ (r=1,\ldots,
d)$ and $k_r$ is called rate constant of the $r$-th
elementary reaction. In the latter case (23) represents an
autonomous system with polynomial right sides.

Let $L$ be a linear envelope of the family $\{\gamma_r\}_{r=1}^d$.
If $L \ne \Bbb{R}^n$, then there exist such $a_i \  (i=1, \ldots ,n)$,
not all equal to sero, that for all $r=1,\ldots ,d$ the equalities
$$
\sum_{i=1}^n a_i \gamma_{ri}=0
$$
are satisfied, from which for system (23) follows that
$$
\sum_{i=1}^n a_i c_i(t)={\rm const} \eqno {(24)}
$$

Relationships (24) are called stoichiometric conservation laws. If
all $a_i$ are positive, then the corresponding stoichiometric law is
called {\it the positive conservation law} [15]. In MAL positive
conservation laws takes place rather often (but not always).

As it is known [15], balance polyhedrons are intersections of
affine subspaces of the form $(L+c)$, where $c$ is a constant
vector, with a cone of non-negative vectors (first orthant)
in $\Bbb{R}^n$. Balance polyhedrons represent positive invariant with
respect to (23) convex sets (one can find the proof of their
positive invariance in [15]). If there exists at least one of positive
conservation law, they are compact.

The question arises: under what conditions does the norm exist in $\Bbb{R}^n$ according
to which the system (23) is contracting in all balance polyhedrons and independent of
rate constants?

{\bf Definition 8}. Mechanism (22) is called {\it dissipative}, if for system (23) there
exists a norm, contracting in all balance polyhedrons irrespective of rate constants (in
other words, the contracting norm depends on the mechanism only).

We use the notation $M_{ri}$ for the operator in $\Bbb{R}^n$,
represented by the matrix, in the $i$-th column of which there are
components of the vector $\gamma_r$, and on other places -- zeros.
The subspace $L$ is invariant with respect to all $M_{ri}$ [15].
The notation $M'_{ri}$ stays for restriction of $M_{ri}$ on $L$.

{\bf Theorem 8}. Let for mechanism (22) exist at least one
positive conservation low. This mechanism is dissipative if and only if
the family $\{M'_{ri}: \ \alpha_{ri} > 0\}$ is simultaneously dissipative.

{\bf Proof}. {\it Sufficiency}. It is known [15] that the Jacobi matrix $J_c$ of
system (22) at the point $c$, whose coordinates are positive, has
the form
$$
J_c=\sum_{\alpha_{ri}>0} \alpha_{ri}\frac{w_r}{c_i}M_{ri}.
\eqno {(25)}
$$

Matrices $J_c$ belong to the convex cone produced by the family $\{M_{ri} \vert \
\alpha_{ri} > 0\}$. Besides, the difference of any two solutions (23) from one balance
polyhedron belongs to the subspace $L$. Under the conditions of lemma 3 and theorem 2
obtain the existence of contracting norm in the subspace $L$. It can be expanded onto all
$\Bbb{R}^n$.

{\it Necessity}. Matrices $M_{ri} \ (\alpha_{ri}>0)$ belong to the closure of
the family of matrices $J_c$ for arbitrary non-negative vectors $c$
and rate constants $k_r$. To prove this, first let consider the case when
$c_j\  (j=1,\ldots ,n)$
and $k_r$ are fixed and all $k_l (l \ne r)$ tend to zero. In the limit in
(25) only the sum for given $r$ is left. Further on,
fix all $c_j>0\  (j\ne i)$ and let $c_i$ tend to zero,
changing $k_r$ so that the equality $\alpha_{ri}w_r/c_i=1$
holds true. Then all the terms except one tend to zero and
in the limit we obtain $M_{ri}$.

Thus, the matrices $M'_{ri}\  (\alpha_{ri}>0)$ belong to the
closure of the family of restrictions of the matrices $J_c$ on the
subspace $L$. Hence, according to lemma 3 and theorem 2 the necessity
follows. The theorem is proved.

Note that matrices $M_{ri}$ represent matrix-columns (in each
matrix there is only one non-zero column) and that means that the rank of
each of them is equal to unity. We come to the problem of
simultaneous dissipativity of the finite family of operators of rank
1.

Note that dissipative mechanisms of reactions were studied in
details in [12]. In particular, some classes of dissipative
mechanisms are pointed out and all dissipative mechanisms for $n=3, \
\sum_{i=1}^3 \alpha_{ri} \leq 3,\  \sum_{i=1}^3 \beta_{ri}\leq 3
\  (r=1,\ldots ,d),\  c_1+c_2+c_3={\rm const}$ enumerated.

In the next subsection are obtained necessary and
sufficient conditions of simultaneous dissipativity of the operators
of rank 1 in $\Bbb{R}^2$ (corresponding to the case $dim L=2$) and some
sufficient conditions of simultaneous dissipativity of
matrix-columns.

\subsection{ Constructive Conditions of Simultaneous Dissipativity of
\hfill \protect\linebreak One-Dimensional Operators }

Before consideration of simultaneous dissipativity
of operators of rank 1, find out what can be said about
dissipativity of one such operator. From necessary and sufficient
conditions (see the paragraph after lemma 6) follows that the norm
in which the given operator of rank 1 is dissipative exists
if and only if it has a negative eigenvalue.

Positive semi-trajectories of system (3) corresponding to the
initial condition $x(0)=x_0$ are in this case rectilinear segments
parallel to the image of $A$ and connecting $x_0$ with $Ker\ A$.
Operator A of rank 1 is dissipative in the
given norm if and only if for any point $x\ (\Vert x
\Vert=1)$ there exists such $\varepsilon > 0$ that $\Vert x+\varepsilon
Ax\Vert \leq 1$. It means that the negative number belongs
to the spectrum of $A$, and the image of $A$ is orthogonal to its
kernel (in the given norm, the subspace $E_2$ is orthogonal to
$E_1$ if $\Vert x+y \Vert \geq \Vert x \Vert$ for any $x\in E_1, y
\in E_2$ [7]).

Let now be given a family $\{M_k\}_{k=1}^m$ of operators of rank 1
in $\Bbb{R}^n$. Each of them can be represented in the form
$(\cdot \ ; \psi_k) \varphi_k$, i.e. $M_kx=(x\ ; \psi_k)\varphi_k$
where $(\cdot \ ; \cdot )$ is the standard scalar product in
$\Bbb{R}^n$. The vectors
$\varphi_k$ and $\psi_k$ are determined by the operator $M_k$
unambiguously (up to scalar factors). Let $\lambda_k=(\varphi_k;
\psi_k)$, i.e. $\lambda_k$ is an eigenvalue of $M_k$ (either it is
the only non-zero eigenvalue, or $0$, if the operator $M_k$ is
nilpotent). As it has already been mentioned, for simultaneous
dissipativity of $\{M_k\}$ the conditions
$$
\lambda_k < 0\  (k=1,\ldots ,m) \eqno {(26)}
$$
are necessary.

Assign to each operator $M_k$ the projector $P_k$ projecting parallel to the image of
$M_k$ on the kernel of $M_k$. It is easy to see that $P_k=I-M_k/\lambda_k$. By virtue of
the above-mentioned condition of dissipativity of the operator of rank 1 in the given
norm the operator $M_k$ is dissipative in some norm if and only if $P_k$ is contraction
in this norm.

All $P_k$ can be contractions in one norm if and only if all products of the form
$\prod_{j=1}^q P_{k_j}$ ($q \in N$ is arbitrary; $k_j \in \{1, \ldots ,m\}$ and they are
not necessarily different) are jointly bounded. We come to the following conclusion.

{\bf Lemma 7}. The family $\{M_k\}_{k=1}^m$ of operators of rank 1 is simultaneously
dissipative if and only if the conditions (26) are satisfied and all products of the form
$\prod_{j=1}^q P_{k_j}\  (q\in N$ is arbitrary; $k_j \in \{1,\ldots , m\}$ and they are
not necessarily different) are jointly bounded. As a contracting norm one can take
$$
\Vert x \Vert =\sup_{q\in N,1\leq\ k_j\leq m}\{\Vert x \Vert_0,
\bigl\| (\prod_{j=1}^q P_{k_j})x\bigr\|_0\}, \eqno {(27)}
$$
where $\Vert . \Vert_0$ is any norm in $\Bbb{R}^n$.

{\bf Proof}. All statements of the lemma, exept the latter, follow
immediatily from the above reasonings. Further on, if all products
$\prod_{j=1}^q P_{k_j}$ are jointly bounded, then
$$
\sup_{q\in N,1\leq\ k_j\leq m}\{\Vert x \Vert_0,
\bigl\| (\prod_{j=1}^q P_{k_j})x\bigr\|_0\} < {\infty}
$$
for each $x \in \Bbb{R}^n$. This expression possesses all properties of norm and all
operators $P_k$ are contractions in such norm, i.e. all $M_k$ are simultaneously
dissipative. The lemma is proved.

From lemma 7 follows a simple consequence.

{\bf Corollary 1}. If all $\varphi_k$ are collinear (images of $M_k$ coincide) or all
$\psi_k$ are collinear (kernels of $M_k$ coincide) and $(\varphi_k ; \psi_k) < 0$ for all
$k=1,\ldots,m$, then the operators $M_k \ (k=1,\ldots,m)$ are simultaneously dissipative.
As corresponding contracting norm one can take
$$
\sup_{q\in N,1\leq k_j\leq m} \{\Vert x \Vert_0, \Vert
P_kx\Vert_0\}.
$$
To demonstrate this, it is sufficient to note that in these cases
$$
\prod_{j=1}^q P_{k_j}=P_{k_1}$$ or $$ \  \prod_{j=1}^q P_{k_j}=P_{k_q},
$$
respectively.

{\bf Remark 10}. If not all $\varphi_k$ are collinear, then as a norm in
lemma 7 one can take
$$
\sup_{q\in N, 1\leq k_j\leq m}\bigl\| (\prod_{j=1}^q P_{k_j})x\bigr\|_0.
\eqno {(28)}
$$

The criterion established in lemma 7 is not constructive.
Constructive criteria of simultaneous dissipativity of finite
family of operators of rank 1 in $\Bbb{R}^n$ have been obtained
only at $n=2$ (for arbitrary $n$ there exist sufficient conditions
for one class of operators; they are given at the end of the
section). Pass to the consideration of the case $n=2$.

Consider the family $\{M_k\}_{k=1}^m$ of the operators of rank 1 in
$\Bbb{R}^2$. As before, represent each operator $M_k$ in the form
$(\cdot \ ; \psi_k) \varphi_k$. Let first $m=2$.

{\bf Lemma 8}. The operators $M_1=(\cdot \ ;\psi_1)\varphi_1$ and
$M_2=(\cdot \ ;\psi_2)\varphi_2$ are simultaneously dissipative in
$\Bbb{R}^2$ if and only if the condition
$$
\biggl| \frac {(\varphi_1;\psi_2)\ \cdot \ (\varphi_2;\psi_1)}
{(\varphi_1;\psi_1)\cdot (\varphi_2;\psi_2)}\biggr| \leq 1 \eqno {(29)}
$$
is satisfied together with the conditions
$$
(\varphi_1; \psi_1) < 0;\  (\varphi_2; \psi_2) < 0.
$$
As a corresponding contracting norm one can take
$$
\Vert x\Vert = \max \{\Vert x\Vert_0,\Vert P_1x\Vert_0,\Vert
P_2x\Vert_0,\Vert P_1P_2x\Vert_0,\Vert P_2P_1x\Vert_0\}.
\eqno {(30)}
$$

{\bf Proof}. In $\Bbb{R}^2$ the projectors $P_1$ and $P_2$ have
rank 1 and are represented in the form
$$
P_1=(\cdot \ ; \eta_1)\chi_1;\  P_2=(\cdot \ ;\eta_2)\chi_2,
$$
where $\eta_1, \eta_2, \chi_1, \chi_2$ are some vectors in $\Bbb{R}^2$.

The operators $\prod_{j=1}^q P_{k_j}$ are bounded
when the spectrum of the operator
$(P_1P_2)$ lies on the segment $[-1;1]$:
$$
\vert (\chi_1; \eta_2)\cdot (\chi_2; \eta_1)\vert \leq 1. \eqno{(31)}
$$

In a standard orthonormalized basis $P_k$ acts like this:
$$
P_kx=\frac {1}{(\varphi_k;\psi_k)}\cdot
\left(
\left(
\begin{array}{r}
x^{(1)} \\[2mm]
x^{(2)}
\end{array}
\right);
\left(
\begin{array}{r}
\varphi_k^{(2)}\\[2mm]
-\varphi_k^{(1)}
\end{array}
\right)
\right)
\cdot
\left(
\begin{array}{r}
\psi_k^{(2)} \\[2mm]
-\psi_k^{(1)}
\end{array}
\right)
$$
where
$$
\left(
\begin{array}{r}
a^{(1)} \\[2mm]
a^{(2)}
\end{array}
\right)
$$
denotes the vector with the coordinates $a^{(1)}$ and $a^{(2)}$. Hence
$$
(\eta_1; \chi_2)=\frac {(\varphi_1;\psi_2)}{(\varphi_2;\psi_2)};
$$
$$
(\chi_1;\eta_2)=\frac{(\varphi_2;\psi_1)}{(\varphi_1;\psi_1)},
$$
i.e. condition (31) takes the form (29).

To complete the proof, use lemma 7. To
check a possibility of choosing corresponding norm in the form (30),
note that
$$
(P_1P_2)^rP_1=(\eta_1;\chi_2)^r\cdot (\eta_2;\chi_1)^r\cdot P_1;
$$
$$
(P_2P_1)^rP_2=(\eta_1;\chi_2)^r\cdot (\eta_2;\chi_1)^r\cdot P_2
$$
for any $r \in N$. It means that with the account of (31), in (27)
one can restrict oneself to finite number of products. The lemma
is proved.

{\bf Remark 11}. If $\varphi_1$ and $\varphi_2$ are non-collinear, then as
required norm we can take
$$
\max \{\Vert P_1x\Vert_0, \Vert P_2x\Vert_0, \Vert P_1P_2x\Vert_0,
\Vert P_2P_1x\Vert_0\}.
$$

This follows from remark 10. Then the ball of the norm is
determined by the inequalities
$$
\vert(x;\eta_1)\vert \leq \min \biggl\{\frac {1}{\Vert\chi_1\Vert_0},
\frac {1}{\vert(\chi_1;\eta_2)\vert \cdot \Vert \chi_2\Vert_0}\biggr\};
$$
$$
\vert(x;\eta_2)\vert \leq \min \biggl\{\frac {1}{\Vert \chi_2\Vert_0},
\frac {1}{\vert (\chi_2;\eta_1)\vert \cdot \Vert\chi_1\Vert_0}\biggr\},
$$
i.e. it is parallelogram.

Also note that for simultaneous
dissipativity of a family the dissipativity of each operator
from convex envelope of the family is insufficient. To see this,
consider the operators represented by the matrices
$$
M_1=
\left(
\begin{array}{rr}
-1 & 1 \\[2mm]
 0 & 0
\end{array}
\right);\
M_2=
\left(
\begin{array}{rr}
 0 & 0 \\[2mm]
-2 &-1
\end{array}
\right).
$$

Each of them is dissipative in its norm. It is easy to show that
spectrum of any non-trivial convex combination of $M_1$ and $M_2$
lies in open left half-plane. Nevertheless
$$
\frac {(\varphi_1 ; \psi_2) \cdot (\varphi_2 ; \psi_1)}
 {(\varphi_1 ; \psi_1) \cdot (\varphi_2 ; \psi_2)} =-2,
$$
i.e. condition (29) is not satisfied.

Reasoning like in proof of
lemma 8, it is easy to obtain a criterion of a
simultaneous dissipativity for arbitrary $m$. The result is a set
of conditions of the form
$$
(\varphi_k ; \psi_k) < 0 \  (k=1, \ldots , m); \eqno {(32)}
$$
$$
\biggl| \frac
{(\varphi_{k_1};\psi_{k_2}) \cdot (\varphi_{k_2};\psi_{k_3}) \cdot
\ldots \cdot (\varphi_{k_q};\psi_{k_1})}
{(\varphi_{k_1};\psi_{k_1}) \cdot (\varphi_{k_2};\psi_{k_2}) \cdot
\ldots \cdot (\varphi_{k_q};\psi_{k_q})}
\biggr| \leq 1, \eqno {(33)}
$$
where $\{k_j\}_{j=1}^q$ is a set of different numbers from
1 to $m$, and inequalities (33) holds for all such sets. The
number of conditions has the order $O((m-1)!)$ and for any large $m$
testing of these conditions becomes unrealizable. It turns out,
however, that among inequalities (33) there are dependent ones and
the number of conditions can be reduced.

{\bf Theorem 9}. Let the vectors $\psi_k \ (k=1,\ldots ,m)$ lie in one
half-plane clockwise. Then the family of operators
$\{M_k\}_{k=1}^m$ where $M_k=( \cdot ; \psi_k) \varphi_k $ is
simultaneously dissipative if and only if the
vectors $\varphi_k \ (k=1,\ldots ,m)$ lie in one half-plane clockwise
and the conditions (32) and the followings ((34), (35)) are satisfied:
$$
\biggl| \frac
{(\varphi_k ; \psi_{k+1}) \cdot (\varphi_{k+1} ; \psi_k)}
{(\varphi_k ; \psi_{k}) \cdot (\varphi_{k+1} ; \psi_{k+1})}
\biggr| \leq 1 \eqno {(34)}
$$
$$
(k=1, \ldots , m \  \mbox{with} \  \varphi_{m+1}=-\varphi_1;
\psi_{m+1}=-\psi_1);
$$
$$
\left\{
\begin{array}{rcl} \displaystyle
\biggl| \frac
{(\varphi_1 ;\psi_2) \cdot (\varphi_2 ;\psi_3) \cdot
\ldots \cdot (\varphi_m ;\psi_1)}
{(\varphi_1 ;\psi_1) \cdot (\varphi_2 ;\psi_2) \cdot
\ldots \cdot (\varphi_m ;\psi_m)}
\biggr| & \leq & 1; \\[4mm]
\displaystyle \biggl| \frac
{(\varphi_1 ;\psi_m) \cdot (\varphi_m ;\psi_{m-1}) \cdot
\ldots \cdot (\varphi_2 ;\psi_1)}
{(\varphi_1 ;\psi_1) \cdot (\varphi_2 ;\psi_2) \cdot
\ldots \cdot (\varphi_m ;\psi_m)}
\biggr| & \leq & 1.
\end{array}
\right. \eqno {(35)}
$$
The corresponding norm can be chosen polyhedral (a norm, whose
ball is polygon).

{\bf Proof}. {\it Necessity}. Let $l_k$ be the kernels of the operators $M_k$
(i.e. straight lines orthogonal to $\psi_k$). Straight lines $l_k$
divide the plane into $2m$ sectors. If among the vectors $\psi_k$
there are
collinear, then some sectors are singular, but this does not change the
further reasonings. In each sector $G$ and for each
$p \in \{1, \ldots ,m\}$
$$
sign (x_1; \psi_p) = sign (x_2; \psi_p)
$$
for all $x_1\in int\ G,\ \ x_2\in int\ G$.

Let $G_r$ be a sector lying between corresponding rays of
straight lines $l_r$ and $l_{r+1}$ (where $l_{m+1}=l_1$). It is
enough to consider the sectors
$\{G_k\}_{k=1}^m$ into which one half-plane is divided, since
for sectors lying in vertical
angles to $G_k$ the reasons are the same.

Note that by inequality (32) for each operator $M_k$
the projector $P_k$ is determined, which operates in each sector $G_k$ as a
projector in the direction $\upsilon_{kr}=sign (x; \psi_k)\cdot \varphi_k
\  (x \in G_r)$ onto the straight line $l_k$.

A norm with respect to which all $M_k$ are
dissipative exists if and only if there exist a convex body $Q$
symmetrical with respect to $0$ and positively invariant with
respect to all systems of the following form
$$
\frac {dx}{dt}=\sum_{k=1}^m h_k(t)(x;\psi_k)\varphi_k, \eqno {(36)}
$$
where $h_k(t)$ is any function piecewise continuous and non-negative for
$t \geq 0.$ The sufficiency is
evident (suppose $h_k(t)\equiv 1, h_j(t)\equiv 0$ for $j \ne k$ and
come to dissipativity of $M_k$ with respect to $Q$). To prove
the necessity, it is sufficient to make an estimation analogous to
that made in the proof of theorem 1:
$$
\frac {d}{dt}\Vert x(t)\Vert_Q = N_Q\bigl(x(t),\sum_{k=1}^m h_k(t)
M_kx(t)\bigr) \leq
$$
$$
\leq \gamma_Q \bigl(\sum_{k=1}^m h_k(t)M_kx(t)\bigr)\cdot
\Vert x(t)\Vert_Q \leq 0.
$$
Here $\Vert . \Vert_Q$ is a norm whose unit ball is $Q$.

Since $(x;\psi_k)\varphi_k=\vert(x;\psi_k)\vert \cdot \upsilon_{kr}$
at $x \in G_r$, then (36) can be rewritten as follows:
$$
\frac {dx}{dt}=\sum_{k=1}^m y_k(t)\upsilon_{kr} \eqno {(37)}
$$
where $y_k(t)$ is piecewise continuous and non-negative for $t \geq
0$. Thus, it is sufficient to construct such a polygon $W$
that from each point of its boundary $\partial W$ all the vectors
$\upsilon_{kr}$ are not directed into the exterior of $W$. Then
one can take
$$
Q=co\ \{W \cup (-W)\}.
$$

Let (37) have at least one unbounded solution, whose positive
semi-trajectory lies inside one of sectors. Then
(36) has an unbounded solution, i.e. the operators $M_k$ are not
simultaneously dissipative.

The notation $C\{\upsilon_{kr}\}$ is used for a convex cone
produced by $\{\upsilon_{kr}\}_{r=1}^m$.

Let this cone concide with $\Bbb{R}^2$ at least in one sector $G_r$ (i.e.
the vectors generating it do not lie in one half-plane). Then as $y_k(t)$
one can choose such constants that
$\upsilon =\sum_{k=1}^m y_k \upsilon_{kr} \in G_r$, and then,
drawing a ray from the point $x_0 \in int G_r$ in the direction of
$\upsilon $, obtain a positive semi-trajectory of unbounded
solution (47) lying inside $G_r$.

Thus, for simultaneous dissipativity of $\{M_k\}$ it is necessary
to satisfy the conditions
$$
C\{\upsilon_{kr}\} \ne \Bbb{R}^2 \  (k=1, \ldots , m). \eqno {(38)}
$$

If $C\{\upsilon_{kr}\}$ in some sector is a half-plane, then it must
contain the vertical angle to $G_r$ - $\hat G_r$
(and thus intersect with $G_r$ only at zero);
otherwise (37) has an unbounded solution. For each sector $G_r$
consider the boundary of the cone $C\{\upsilon_{kr}\}$. It consists
of two directions. Show that for $G_j$ it is $\upsilon_{jj},
\upsilon_{(j+1),j}$. It is sufficient to show that for $j=1$.

Let $\upsilon_{1,1}$ and $\upsilon_{2,1}$ be collinear and
oppositely directed. Then to satisfy (38) it is necessary tha the
other  $\upsilon_{k1}$ lie on one side of the straight line,
stretched on $\upsilon_{1,1}$. But if $\upsilon_{1,1}$ and
$\upsilon_{2,1}$ are non-collinear, then all other $\upsilon_{k1}$ can
be expanded in terms of the basis $\upsilon_{1,1}, \upsilon_{2,1}$.

Let, for example,
$\upsilon_{3,1}=c_1\upsilon_{1,1}+c_2\upsilon_{2,1},$ and
$\upsilon_{3,1}$ be collinear to one of the basis vectors (for
example, $\upsilon_{1,1}$; the case with $\upsilon_{2,1}$ is considered
analogously). Then $c_2=0$. If $c_1 > 0$ then $\upsilon_{3,1} \in
C\{\upsilon_{1,1}, \upsilon_{2,1}\}$. Let $c_1 < 0 $. Then to satisfy
(38) in $G_1$ it is necessary for $\upsilon_{1,1}$ and
$\upsilon_{3,1}$ to be boundary directions in $C\{\upsilon_{k1}\}$.
Since $\upsilon_{k,(l+1)}=\upsilon_{kl}$, if $k \ne l+1$, and
$\upsilon_{(l+1),(l+1)}=-\upsilon_{l,(l+1)}$, then the same
directions are boundary for $C\{\upsilon_{k2}\}$ as well,
otherwise (38) is not satisfied in $G_2$. Simultaneously
$\upsilon_{2,1} \in C\{\upsilon_{k1}\},\  \upsilon_{2,2}=
-\upsilon_{2,1} \in C\{\upsilon_{k2}\}$. It means
$C\{\upsilon_{k1}\}$ and $\{\upsilon_{k2}\}$ represent half-plane
whose join is all $\Bbb{R}^2$, what is impossible. That means
$c_1 > 0$.

Let now $\upsilon_{3,1}$ be non-collinear neither to
$\upsilon_{1,1}$ nor to $\upsilon_{2,1}$. If $c_1 <0, c_2 <0$, then
in $G_1$ (38) is not satisfied. If $c_1 <0, c_2 >0$, then in $G_2$
there $\upsilon_{3,2}=c_1\upsilon_{1,2}+(-c_2)\upsilon_{2,2}$, i.e again
(38) is not satisfied. Analogous reasonings hold for the case $c_1 >0,
c_2 < 0$, i.e. the only possible case is $c_1 \geq 0, \ c_2 \geq 0$ and
therefore $\upsilon_{3,1} \in C\{\upsilon_{1,1}, \upsilon_{2,1}\}$
(where $C\{x, y\}$ is a convex cone, stretched on the vectors $x$
and $y$).

The case is left when the directions $\upsilon_{1,1}$ and
$\upsilon_{2,1}$ coincide.

Without loss of generality one can assume non-collinearity of
$\upsilon_{3,1}$ and $\upsilon_{1,1}$. Then
$\upsilon_{2,2}$ and $\upsilon_{3,2}$ are boundary directions in
$C\{\upsilon_{k2}\}$. Consequently, $\upsilon_{1,1} \in
C\{-\upsilon_{1,1}, \upsilon_{3,1}\}$ i.e. the directions
$\upsilon_{3,1}$ and $\upsilon_{1,1}$ coincide contrarily to the
assumption. It means that if $\upsilon_{rr}$ and $\upsilon_{(r+1),r}$
are co-directed, all $\upsilon_{kr}$ are collinear, i.e. all $\varphi_k$
are collinear. In this case the directions
$\upsilon_{rr}$ and $\upsilon_{(r+1),r}$
are also boundary.

We call the obtained fact {\it the boundariness condition}.

Since all $\psi_k$ lie clockwise in one half-plane, then it is
easy to check that in sector $G_m$ either all $(x ; \psi_k) \geq 0$
for all $k$ or $(x ; \psi_k) \leq 0$ for all $k$. Thus, by virtue of
(32), all $\varphi_k$ lie in one half-plane. From the boundariness
condition follows that
$\varphi_k \in C\{\varphi_{k-1}, \varphi_{k+1}\}$, i.e vectors
$\varphi_k$ are arranged either clockwise, or anti-clockwise.

Let, for example, $\upsilon_{1,1}=\varphi_1$
(the case $\upsilon_{1,1}=-\varphi_1$ is considered analogously).
Then $\upsilon_{2,1}=\varphi_2$ lies in the half-plane bounded by the
straight line stretched on $\upsilon_1$ and containing $\hat S_1$.
Therefore the direction from $\varphi_1$ to $\varphi_2$ in the
half-plane containing all $\varphi_k$ is the same as from
$\psi_1$ to $\psi_2$, i.e. clockwise.

The necessity of the other conditions is obvious, since
(34)-(35) is simply a part of conditions (33).

{\it Sufficiency}. Let the family $\{\varphi_k\}_{k=1}^m $ be arranged
clockwise in one half-plane and the conditions (32) and (34)-(35)
be satisfied. Assume that among $\varphi_k$ there are non-collinear
vectors, and among $\psi_k$ there are no collinear ones.

The condition of clockwise arrangement of $\varphi_k$ in
one half-plane means that the angle (counted
from $\varphi_1$ clockwise) between $\varphi_1$ and the vectors
$\varphi_1, \varphi_2, \ldots , \varphi_m,$ \linebreak[1] $\varphi_{m+1}=-\varphi_1$
monotonously increases from $0$ to $\pi $. Taking into
account that the angle between $\varphi_{k1}$ and $(-\varphi_{k2})$
is the angle between $\varphi_{k1}$ and $\varphi_{k2}$, taken with
opposite sign, it is easy to conclude that systems
$\{\upsilon_{kr}\}$ (in each sector) lie in one half-plane and are
arranged clockwise (to avoid exiting from corresponding half-plane
we start counting in sector $G_r$ from $\upsilon_{(r+1),r}$).

From conditions (34) follows that in each sector there is a
"convex configuration", i.e. there is vector $x \in G_r$,
representable in the form
$$
x=-\sum_{k=1}^m c_k \upsilon_{kr},
$$
where all $c_k > 0$.

It means that if from one point $\tilde x \in int\ G_r$ one draws
segments $\bar a$ and $\bar b$ in the directions of $\upsilon_{rr}$
and $\upsilon_{(r+1),r}$ up to the crossing with $l_r$ and $l_{r+1}$,
respectively, then these segments together with the segments
connecting $0$ with the point of crossing $\bar a$ with $l_r$ and
$\bar b$ with $l_{r+1}$, respectively, form a convex polygon (if
$\upsilon_{rr}$ and $\upsilon_{(r+1),r}$ are oppositely directed, it
will be a triangle, and if they are non-collinear -- a quadrangle;
as we have seen before they cannot be co-directed).

Due to the
same orientation of $\{\varphi_k\}$ and $\{\psi_k\}$ all the other
$\upsilon_{kr}$ are directed (from point $\tilde x$) into this
polygon, i.e. for any cone $C\{\upsilon_{kr}\}$ directions on the
straight lines $\upsilon_r$ and $\upsilon_{r+1}$ are boundary.

Fix now the point $x_0 \in l_1\  (x_0 \ne 0)$ on the boundary ray
of sector $G_1$ (actually, one can begin from any straight line
$l_k$; we begin from $l_1$). Due to the boundariness condition
either direction from $x_0$ on $l_2$ goes into sector $G_1$, or
direction from $x_0$ on $l_m$ goes into $\hat G_m$.

If one and only one of these statements is true, continue moving in the
corresponding direction (to the neighboring straight line)
till the direction on the neighboring straight line goes into
the neighboring sector. In other words, move from $l_r$ to
$l_{r+1}$ in the direction parallel to $\varphi_r$, if this
direction goes into sector $G_r$ (or, into $\hat G_{r-1}$,
respectively). As a polygon $W$ mentioned after (37) one should
take a polygon formed by the segments which we moved along, and the
segments of those straight lines on which the movement broke (if
exit on the initial ray did not occur, in our case it is a part of
$l_1$ corresponding to $G_1$, then it is a segment connecting $x_0$
with $0$, and a segment of that straight line on which the movement
broke, connecting the point of breaking with zero; if exit on the
initial ray occured, then it is a segment connecting $x_0$ with the
point of exit).

If both statements are satisfied, then as $W$ one can take a
join of two such polygons formed in moving to both sides from
$x_0$.

This algorithm is easy to check proceeding from boundariness
conditions, "convex configuration", and (35) (the latter condition
means that if exit on the initial ray occured in moving in either
side, then the point of exit is no farther from the beginning of
coordinates than the initial point; in particular, if the point of
exit coincides with the initial point, then the formed polygon can
be taken as $W$). The ball of the sought for norm is a polygon.

If some of $\psi_k$ are collinear, then some sectors $G_k$ are
singular. This, however, does not change the results. The reasonings
are analogous to the case when among $\psi_k$ there are no
collinear vectors. The only difference here is the following: some
straight lines $l_k$ correspond to several directions
$\{\varphi_j\}_{j=k_0}^{k_1}$. Then in constructing $W$ one needs
to move along $\varphi_{k_0}$.

In the case when all $\varphi_k$ (or all $\psi_k$) are collinear
(see corollary 1), all the same one can regard that
$\{\varphi_k\}$ and $\{\psi_k\}$ have the same orientation,
starting from (32).

Conditions (34)-(35) are satisfied in this case. The
norm can be chosen polyhedral, if one chooses a polyhedral norm as
$\Vert . \Vert _0$ in (30). The theorem is proved.

{\bf Remark 12}. One can obtain the arrangement of vectors $\psi_k$
required by the conditions of theorem 9 by renumbering vectors
and (if it is necesary) changing signs of some of them.

Thus, the problem of simultaneous dissipativity of a family of
operators of rank 1 in $\Bbb{R}^2$ is solved completely. The number of
conditions to be checked now, in contrast to (33), is only of the order
$O(m)$.

With theorem 9 one can study the MAL mechanism on
dissipativity (and, respectively, on the absence of IME). For
example, let the mechanism be
$$
A_1 \to A_2,\  A_1 \to A_3,\  A_2 \to A_1,\  A_2 \to A_3,
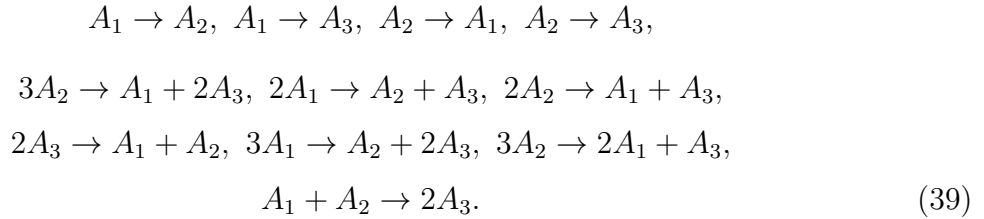
$$
$$
3A_2 \to A_1+2A_3,\  2A_1 \to A_2+A_3,\  2A_2 \to A_1+A_3,
$$
$$
2A_3 \to A_1+A_2,\  3A_1 \to A_2+2A_3,\  3A_2\to 2A_1+A_3,
$$
$$
A_1+A_2 \to 2A_3. \eqno {(39)}
$$

This mechanism possesses positive conservation law
$c_1+c_2+c_3={\rm const}$. The corresponding subspace is the plane
$$
c_1+c_2+c_3=0.
$$
Obviously, $dimL=2$, and one can use theorem 9.
Writing matrices $M'_{ri}$ and using theorem 9, let make sure that
mechanism (39) is dissipative. The corresponding norm in the
subspace $L$ has the form
$$
\Vert c \Vert=\vert c_1 \vert + \vert c_2 \vert .
$$
It can be expanded onto all $\Bbb{R}^3$, for example, in this way:
$$
\Vert c\Vert=\vert c_1\vert+\vert c_2\vert+\vert c_1+c_2+c_3\vert .
$$

To complete the section, consider the question of simultaneous
dissipativity of the finite family of operators of rank 1 of
special form in $\Bbb{R}^n$ for arbitrary $n$. Namely, we consider
operators represented by matrix-columns. Let obtain sufficient
conditions of simultaneous dissipativity of such operators.

Let the basis $\{e_k\}_{k=1}^n $ and the norm
$$
\Vert x\Vert =\sum_{k=1}^n p_k\vert x_k\vert \eqno {(40)}
$$
be given in $\Bbb{R}^n$, where $p_k > 0\  (k=1,\ldots , n),\ x_k\ $
is the $k$-th coordinate of vector $x$ in the basis $\{e_k\}$.
Norm (40) coincides with $l^1$ norm with recpect to the basis
$\{e_k/p_k\}$. Therefore, the necessary and sufficient dissipativity
conditions of the operator $A$ represented by the matrix
$(a_{ij})_{i,j=1}^n$ according to remark 9 have the form
$$
p_ia_{ii}+\sum_{j \ne i} p_j\vert a_{ji}\vert \leq 0 \
(i=1,\ldots ,n). \eqno {(41)}
$$

Let now there be a family of operators, represented by the
matrix-columns $A_{kl_k}$ \linebreak[1]
$(k=1,\ldots,n;\ l_k=0,\ldots ,r_k),$
where $A_{kl_k}$ is the $l_k$-th matrix with non-zero $k$-th
column:
$$
A_{kl_k}=
\left(
\begin{array}{ccccccc}
0 & \ldots & 0 & a_{1k}^{(l_k)} & 0 & \ldots & 0 \\[4mm]
0 & \ldots & 0 & a_{2k}^{(l_k)} & 0 & \ldots & 0 \\[4mm]
 & \ldots & & \ldots & & \ldots & \\[4mm]
0 & \ldots & 0 & a_{nk}^{(l_k)} & 0 & \ldots & 0
\end{array}
\right). \eqno {(42)}
$$

Coming from (41), write dissipativity conditions of all operators
in norm (40) with some constants $p_k$:
$$
p_ka_{kk}^{(l_k)}+\sum_{j \ne k} p_j \vert a_{jk}^{(l_k)}\vert \leq
0 \  (k=1,\ldots ,n;\ l_k=0,\ldots ,r_k). \eqno {(43)}
$$

{\bf Theorem 10}. If the system of linear inequalities (43) complemented
by the inequalities
$$
p_k > 0 \  (k=1, \ldots , n) \eqno {(44)}
$$
has a solution, then the family of operators represented by matrices
(42) is simultaneously dissipative.

{\bf Proof}. Solvability of the systems (43)-(44) means the existence of
positive constants $p_k\  (k=1,\ldots ,n)$ for which inequalities
(43) are satisfied, and that is dissipativity condition of all operators
of the family in norm (40). Obviously, in this case the family is
dissipative. The theorem is proved.

Thus, for simultaneous dissipativity of finite family of
operators represented by matrices-columns the solvability of above
written finite system of linear inequalities proves to be
sufficient. To check solvability, one can use algorithms of linear
programming [16].

{\bf Remark 13}. The solution of the system (43)-(44) exists if there exists
solution of the system of $(n-d)$ linear inequalities
complemented by inequalities (44) (where $d$
is the number of those $k$ for which $r_k=0$; evidently
$0 \leq d \leq n-1$). To prove this, assume
$$
a_{kk}=\max_{0 \leq l_k \leq r_k} a_{kk} ^{(l_k)}; \
a_{jk}=\max_{0 \leq l_k \leq r_k} \vert a_{jk}^{(l_k)}\vert \ (j \ne k)
$$
$$
(k=1,\ldots ,n).
$$

Consider the system
$$
p_ka_{kk}+\sum_{j \ne k}p_ja_{jk}\leq 0 \  (k=1,\ldots ,n).
\eqno {(45)}
$$

Obviously, if the set $\{p_k\}$ satisfies the system (44)-(45), then
it satisfies the system (43)-(44) as well. Numbers $k$ for which
$r_k=0$ are excluded. Therefore, in system (45) there are
$(n-d)$ inequalities.

{\bf Remark 14}. For $n=2$ theorem 10 provides necessary and sufficient
conditions of simultaneous dissipativity. To demonstrate that, note that
for operator $M_k$ of the considered form the vector $\psi_k$
(see the notation at the beginning of the subsection) is directed
along one of the coordinate axes. Therefore (see the proof of
sufficiency in theorem 9), if the family is simultaneously
dissipative, then one can choose parallelogram as a ball of the
corresponding norm, with vertices on coordinate axes, i.e. the norm
is of the form (40). In the case of
arbitrary $n$ the conditions of theorem 10 are already not
necessary. To see this, let
$$
A_1=
\left(
\begin{array}{rrr}
-1 & 0 & 0 \\[4mm]
-1 & 0 & 0 \\[4mm]
 1 & 0 & 0
\end{array}
\right);\
A_2=
\left(
\begin{array}{rrr}
 0 &-1 & 0 \\[4mm]
 0 &-1 & 0 \\[4mm]
 0 & 1 & 0
\end{array}
\right).
$$

The system of linear inequalities
$$
\left\{
\begin{array}{rcl}
-p_1+p_2+p_3 & \leq & 0; \\[4mm]
 p_1-p_2+p_3 & \leq & 0
\end{array}
\right.
$$
has no positive solutions. Nevertheless, simultaneous
dissipativity exists, since each of the operators is dissipative in
its norm and $\varphi_1=\varphi_2$ (see corollary 1).\\[6mm]
\begin{center}
{\bf \Large
Conclusion} \\ [4mm]
\end {center}

Let us resume. The {\it infinitesimal Moore effect} (IME) in an interval space for a
smooth autonomous system on a positively invariant convex compact is studied. The local
conditions of the absence of IME in terms of Jacobi matrices field of the system are
obtained. The relation between the absence of IME and simultaneous dissipativity of the
Jacobi matrices is established, and some sufficient conditions of simultaneous
dissipativity are obtained.

On the basis of the conducted analysis the reason of weak efficiency of interval stepwise
methods is pointed out. The main reason is that to solve the problem of absence of IME in
the system and to construct corresponding interval space one needs analysis of
simultaneous dissipativity of Jacobi matrices of system and constructing a contracting
norm. The latter questions are rarely solved constructively. Besides, in sufficiently
rich interval spaces (for example, in using standard intervals -- rectangular
parallelepipeds) IME is almost always present. One should, however, remember that the
notion of the Moore effect in the work is treated sufficiently strongly. The final
conclusion on the efficiency of stepwise interval methods can be drawn only after
studying {\it asymptotic Moore effect} (AME). It should also be noted that there may be
definitions of interval spaces, different from definition 1.

Some particular classes of systems without IME and corresponding
interval spaces are pointed out. These results can be used in
solving by interval methods particular systems from the
pointed out classes.

\newpage
\begin{center}
{\bf \Large
References}\\ [4mm]
\end{center}

1.~Moore R.E. {\em Interval analysis}.-- N.-Y.: Prentice-Hall, 1966.

2.~Kalmykov S.A., Shokin Yu.I., Yuldashev Z.Kh. {\em Methods of interval analysis}.
Novosibirsk: Nauka, 1986. 221 pp.

3.~Chernousko F.L. {\em Optimal guaranteed estimations of uncertainty by means of
ellipsoides}, Izv. AN SSSR. Tekhnich. kibernetika. 1980. N 5. PP.5--10.

4.~Kracht M., Schoeder G. {\em Zur Intervallrechnung in linear Raumen.-- Computing}.
1973. V.11. PP.73--79.

5.~Ratschek H. {\em Nichtnumerische Aspecte der Intervallarithmetik}, Interval
Ma\-thematics. Berlin-Heidelberg: Springer-Verl., 1975. PP.48--74.

6.~Kuratovskii K. {\em Topology}. V.1. Moscow: Mir, 1966.  594 pp.

7.~Belitskii G.R., Lyubich Yu.I. {\em Matrix norms and their applications}. Kiev: Naukova
dumka, 1984. 151 pp.

8.~Kantorovich L.V., Akilov G.P. {\em Functional analysis}. Moscow: Nauka, 1977. 706 pp.

9.~Lozinsky S.M. {\em Error estimation of numerical integration of ordinary differential
equations}, Izv. vuzov. Ser. Math. 1958. N.5. PP.52--90.

10.~Bylov B.F., Vinograd P.A., Grobman D.M., Nemytskii V.V. {\em Lyapunov exponent theory
and its applications to the problems of stability}. Moscow: Nauka, 1966. 576~pp.

11.~Verbitskii V.I., Gorban A.N. {\em Simultaneously dissipative operators and their
applications in dynamical systems}. Krasnoyarsk, 1987. 32 pp. (Preprint /AS USSR, SB,
Computing Center).

12.~Verbitskii V.I., Gorban A.N. {\em Thermodynamical restrictions and
quasi-termo\-dynamicity conditions in chemical kinetics}. In: Mathematical problems of
chemical kinetics / Ed. by K.I.Zamaraev and G.S.Yablonskii. Novosibirsk: Nauka, 1989.
PP.42--83.

13.~Burbaki N. {\em Lie groups and algebras}. Moscow: Mir, 1976. 496 pp.

14.~Volpert A.I., Khudyaev S.I. {\em Analysis in the classes of discontinuous functions
and equations of mathematical physics}. Moscow: Nauka, 1975. 394 pp.

15.~Gorban A.N., Bykov V.I., Yablonskii G.S. {\em Essays on chemical relaxation}.
Novosibirsk: Nauka, 1986. 300 pp.

16.~Golshtein Ye.G., Yudin D.B. {\em New tendencies in linear programming }. Mos\-cow:
Sov.radio, 1966. 524 pp.

17.~Verbitskii V.I., Gorban A.N., Utjubaev G.Sh., Shokin Yu.I. {\em Moore effect in
interval spaces}, Dokl. AN SSSR. 1989. V. 304, N 1. PP.17--21.

18.~Bykov V.I., Verbitskii V.I., Gorban A.N. {\em On one estimation of solution of Cauchy
problem with uncertainty in initial data and right part}, Izv. vuzov, Ser. Math. 1991. N.
12. PP.5--8.

19.~Verbitskii V.I., Gorban A.N. {\em Simultaneously dissipative operators and their
applications}, Sib. Math. Journal. 1992. V.33, N 1. PP.26--31.

20.~Verbitskii V.I., Gorban A.N. {\em Simultaneously dissipative operators and
quasi--thermodynamicity of the chemical reactions systems}, Advances in Modelling and
Simulation, Tassin (France): AMSE Press. 1991. V.26, N 1. PP.13--21.

21.~Verbitskii V.I., Gorban A.N. {\em On one approach to the analysis of stability of
nonlinear systems and differential inclusions}, Advances in Modelling and Analysis, A.
Tassin (France): AMSE Press. V.19, N 4, 1994. PP.15--27

22.~Verbitskii V.I., Gorban A.N. {\em Stability analysis and solution evaluation for
nonlinear systems by ``Jacobian fields" and Liapunov norms}. AMSE Transactions,
Scientific Siberian, A, V. 4. Dynamics. Tassin (France): AMSE Press. 1992. PP.104--133.

\end{document}